\documentclass{aastex61}
\pdfoutput=1

\usepackage{amsmath}
\usepackage{calc}
\providecommand{\keywords}[1]{\textbf{Keywords ---} #1}

\definecolor{Red}{RGB}{255,37,21}
\definecolor{Green}{RGB}{81,153,74}

\providecommand{\bb}[1]{\mathbf{#1}}

\newcommand{\beq}{\begin{equation}}
\newcommand{\eeq}{\end{equation}}
\newcommand{\bqa}{\begin{eqnarray}}
\newcommand{\eqa}{\end{eqnarray}}

\begin{document}
\title{A GPU implementation of the Correlation Technique for Real-time Fourier Domain Pulsar Acceleration Searches.}

\author[0000-0002-0967-1332]{Sofia Dimoudi}
\altaffiliation{Currently at the Centre for Advanced Instrumentation (CfAI), Physics Department, Durham University, Durham DH1 3LE, UK}
\affiliation{Oxford e-Research Centre, Department of Engineering Science, University of Oxford, 7 Keble Road, Oxford OX1 3QG, UK}

\author[0000-0003-2797-0595]{Karel Adamek}
\affiliation{Oxford e-Research Centre, Department of Engineering Science, University of Oxford, 7 Keble Road, Oxford OX1 3QG, UK}

\author{Prabu Thiagaraj}
\altaffiliation{Currently at the Raman Research Institute (RRI), India}
\affiliation{Jodrell Bank Centre for Astrophysics, The University of Manchester, Macclesfield, Cheshire, SK11 9DL, UK}

\author[0000-0001-5799-9714]{Scott M. Ransom}
\affiliation{National Radio Astronomy Observatory, Charlottesville, VA, United States }

\author{Aris Karastergiou}
\affiliation{Department of Physics, University of Oxford, The Denys Wilkinson Building, Keble Road, Oxford OX1 3RH, UK}

\author[0000-0003-1756-3064]{Wesley Armour} 
\affiliation{Oxford e-Research Centre, Department of Engineering Science, University of Oxford, 7 Keble Road, Oxford OX1 3QG, UK}

\correspondingauthor{Wesley Armour}
\email{wes.armour@oerc.ox.ac.uk}

\begin{abstract}
The study of binary pulsars enables tests of general relativity. Orbital motion in binary systems causes the apparent pulsar spin frequency to drift, reducing the sensitivity of periodicity searches. Acceleration searches are methods that account for the effect of orbital acceleration. Existing methods are currently computationally expensive, and the vast amount of data that will be produced by next generation instruments such as the Square Kilometre Array (SKA) necessitates real-time acceleration searches, which in turn requires the use of High Performance Computing (HPC) platforms. We present our implementation of the Correlation Technique for the Fourier Domain Acceleration Search (FDAS) algorithm on Graphics Processor Units (GPUs). The correlation technique is applied as a convolution with multiple Finite Impulse Response filters in the Fourier domain. Two approaches are compared: the first uses the NVIDIA cuFFT library for applying Fast Fourier Transforms (FFTs) on the GPU, and the  second contains a custom FFT implementation in GPU shared memory. We find that the FFT shared memory implementation performs between 1.5 and 3.2 times faster than our cuFFT-based application for smaller but sufficient filter sizes. It is also 4 to 6 times faster than the existing GPU and OpenMP implementations of FDAS. This work is part of the AstroAccelerate project, a many-core accelerated time-domain signal processing library for radio astronomy.
\end{abstract}


\section{Introduction} \label{sec:sec1}
The Square Kilometre Array (SKA) will transform the field of pulsar astrophysics \citep{2015aska.confE..36K}, by allowing for the discovery of a large fraction of the radio pulsar population beaming towards Earth. One particular area of opportunity for advances in fundamental physics, comes from the potential discovery of a new population of pulsars with companion objects, in orbits that allow more precision tests of theories of Gravity \citep{2010Natur.467.1081D,2014IJMPD..2330004K}. Additional binary neutron star systems, where one or both objects are detected as radio pulsars, relativistic millisecond pulsars with white dwarf companions \citep{2013Sci...340..448A,2014APS..APR.K4003R}, hierarchical triple systems \citep{2014Natur.505..520R}, and of course a pulsar orbiting a black hole, are prize targets of the SKA pulsar surveys.

The yield of the SKA surveys will be large due to a combination of the sensitivity of the telescope, a large number of simultaneously searched tied array beams, and significant investment in computer hardware to search the data. Arguments pertaining to the large number of tied array beams (N\(\sim \)1000), the available spectral bandwidth, and the resolution in time and frequency that is required to discover new pulsars, suggest that storage of the data will be extremely costly. Offline processing of raw pulsar data will give way to direct searches of the observed data streams, recording to disk only the raw data associated with carefully chosen pulsar candidates. There are two unavoidable consequences. First, searching for pulsars, and above all, the types of systems described above, must occur in real time. Second, the algorithms used to perform the search should not compromise on sensitivity. 

Searching for periodic signals modulated by binary motion increases the complexity of simple periodicity searches. Among the compensation methods for binary motion that exist currently, linear acceleration searches are particularly favoured for their simplicity and computational practicality while also delivering gains in signal recovery for a significant fraction of binary pulsars. Acceleration searches employ a linear, one dimensional model for orbital motion related to a constant acceleration, which enables the signal recovery via the application of a number of trial acceleration values incorporated in the traditional search. There are currently two main methods used routinely for acceleration searches. The first method, which we refer to as Time Domain Acceleration Search (TDAS), uses time resampling according to several trial acceleration values, followed by a Fast Fourier Transform (FFT) for each trial. The second method is the Fourier Domain Acceleration Search (FDAS), which applies multiple filters, corresponding to the trial acceleration values, in the Fourier domain with the use of short-length FFTs. FDAS is considered generally faster, and well suited for parallelism, compared to the time domain method. 

Employing acceleration searches on the vast amount of observation data that will be produced from the SKA, together with the requirement for real-time processing, poses a computational challenge, which current CPU platforms may not be able to meet. That is especially true within the strict limitations on energy consumption that the operation of these facilities will have. Special purpose parallel co-processors such as Graphics Processor Units (GPUs) are ideal candidates for FDAS as they offer massive parallelism through thousands of computing cores on a single device with very high memory bandwidth and local caches. Their programming model is particularly suited in exploiting the memory locality that the FDAS technique exhibits with its short-length array operations. In addition, high performance FFT  libraries such as the cuFFT\footnote{For the latest description of the cuFFT library, including accuracy and performance information, see the NVIDIA CUDA Documentation website at \url{http://docs.nvidia.com/cuda/cufft/index.html} } are available for GPUs to easily perform very fast Fourier transforms, and the technology is evolving continuously towards energy efficiency with an increasing performance-per-Watt ratio.

Pulsar processing software with acceleration search capabilities can be found currently in a number of scientific software projects. The SIGPROC package \citep{Lorimer-2011}, and the GPU-enabled PEASOUP library \citep{ewanbarr_2014_10178} use the TDAS method. The FDAS is implemented as part of the PRESTO \citep{2011ascl.soft07017R} package. A tested GPU version of the PRESTO acceleration search currently exists\footnote{Developed by Jintao Luo, available at: \url{https://github.com/jintaoluo/presto2_on_gpu}.}\textsuperscript{,}\footnote{A second GPU version also exists, which has been under development at the time of this study. Developed by Chris Laidler, available at: \url{https://github.com/ChrisLaidler/presto}.}. This existing PRESTO GPU implementation is not optimised specifically for the real-time pulsar processing scenarios mentioned above, and  to this extent, we believe that the community would benefit from a tunable library with a GPU FDAS component  aimed specifically at performing real-time processing of radio telescope data. In this work, we  propose a GPU implementation of the FDAS algorithm, as part of AstroAccelerate \citep{2012ASPC..461...33A}, a GPU enabled processing library for time domain radio astronomy data\footnote{The AstroAccelerate source code is an open source package, licensed under GNU General Public License v3.0 and is publicly available on Github at \url{https://github.com/AstroAccelerateOrg/astro-accelerate}. Zenodo DOI: 10.5281/zenodo.1212488}.

In the following, we give details of the techniques required to perform the aforementioned searches, and show how we have tackled this problem using modern high-performance computing hardware and software techniques. The remainder of this document is organised as follows: Section~\ref{sec:sec2} introduces the basic  theoretical background for the two acceleration search methods mentioned above. In~\ref{sec:sec3} we describe the application of FDAS in more detail, with a focus on the functional workflow and computational considerations. Our core work of the GPU implementation of the FDAS algorithm is explained in detail in section~\ref{sec:sec4}, and section~\ref{sec:sec5} presents and discusses experimental results from performance and signal recovery measurements in comparison to existing software. Section~\ref{sec:sec6} demonstrates the speed of the algorithm on the latest GPU hardware based on real-time parameters. Finally we summarise and draw our conclusions in~\ref{sec:sec7}.

\section{Effects of Orbital Motion in Pulsar Binaries and corrective methods} \label{sec:sec2}

Pulsar searches typically detect the peak signal power at the pulsar spin frequency by applying the FFT to a series of input samples that are obtained during an observation interval. It should be noted that there exists an alternative to the FFT method for periodicity searches, the Fast Folding Algorithm (FFA)\citep{1448981}, which has seen renewed interest in recent years and is being increasingly  used in wide area surveys, \citep[see, e.g.,][]{2017MNRAS.468.1994C}, but it's use is currently still limited. This work is based on FFT methods, which are the most commonly used. 

Orbital motion in short period binary pulsars causes the apparent pulsar spin frequency to change during the observation time, as a result of its varying line-of-sight velocity. The effect of this is to spread the signal power over a number of neighbouring Fourier bins, reducing the sensitivity of the FFT to these objects. Acceleration searches \citep{1984ApJ...279..157M,1990Natur.346...42A,1991ApJ...368..504J,1991ApJ...379..295W,1993ispu.conf..372M,1538-4357-546-1-L25,1538-3881-124-3-1788,doi:10.1093/mnras/stt161} are methods to correct partly for this effect by assuming a constant acceleration over a fraction of the binary orbit. For a circular orbit with period \(P_{orb} \), the line-of-sight velocity and acceleration are sinusoidal in time, as the orbiting pulsar moves away and towards the observer. For a small fraction of the orbit, the acceleration can be assumed to be constant. If the observation time \(T_{obs} \lesssim P_{orb} / 10 \) \citep[see, e.g.][]{1991ApJ...368..504J, 0004-637X-589-2-911,2002A&A...384..532J}, then, using the constant acceleration assumption, the frequency change can be approximated with a linear relationship to the initial frequency for the specified integration time, according to the Doppler effect. 

If \(\upsilon(t)\) is the observed radial velocity of a pulsar along the line of sight, then, using the Doppler formula, a time interval \(\tau \) in the pulsar frame can be related to a corresponding interval \(t\) in the observed frame with the transformation:
\begin{equation} \label{eq:21}
\tau(t) = \tau_{0} \left \lbrack 1 + \upsilon(t) / c \right \rbrack ,
\end{equation}
where \(c\) refers to the speed of light,  \(\tau_{0}\) is a constant that is used to maintain the correct sampling during a transformation \citep[e.g.][]{0004-637X-535-2-975} , and terms higher than first order in \((\upsilon / c)\) are neglected. To search for objects with unknown orbital parameters, using Kepler's laws to calculate \(\upsilon(t)\), one would need to search in a five-dimensional parameter space, which would be computationally impractical. Assuming a constant acceleration \
\(\alpha\) during an observation interval  \(T_{obs} \lesssim P_{orb} / 10 \), we can approximate \(\upsilon(t) = \alpha t\), and thus greatly reduce the search requirements. The received signal can then be resampled by running a linear interpolation over the original time series. This time resampling method is used in the time domain acceleration search.

A constant acceleration \(\alpha \) corresponds to a constant frequency derivative, \(\dot{f} \), which is related to the number of Fourier bins, \(z\),  that the signal frequency has drifted during the observation time \(T \), such that:
\begin{equation} \label{eq:22}
\alpha = \frac{\dot{f}}{f_{0}} c = \frac{zc}{f_{0} T^2} \, .
\end{equation}

The initial signal response can then be recovered coherently in the Fourier domain by correcting the Fourier response over a frequency - frequency derivative plane ( \(f - \dot{f} \)), using only local Fourier amplitudes of the FFT of the signal, according to the assumed number of bins the frequency has drifted. Pulsar acceleration searches in the Fourier domain are based on the correlation technique \citep{1538-3881-124-3-1788}, which works by correlating a predicted Fourier response, or template, that corresponds to a frequency derivative, or \(z \) number, with the local Fourier amplitudes along the input FFT.  The process is described mathematically by eq. \ref{eq:23}. The corrected Fourier response \(A_{r_0}\) of the signal at frequency bin \(r_0\), is recovered by correlating the \(m\) bins around \(r_0 \) with the frequency reversed and complex conjugated template \(A^{*}_{r_0 - k}  \) of the predicted normalised Fourier response to a given orbital acceleration.
\begin{equation} \label{eq:23}
A_{r_0} \simeq \sum_{k = [r_0] - m/2}^{[r_0] + m/2} A_{k} A^{*}_{r_0 - k} 
\end{equation}
This is effectively a matched filtering process, which can be efficiently computed in parallel for the whole signal over a number of templates by using Fourier techniques. The correlation technique has certain advantages over the time-domain resampling method \citep{1538-4357-546-1-L25}, and in particular the ability to allow independent and memory local calculations, which is beneficial to the application of parallelism. 

\section{Fourier Domain Acceleration Search Method} \label{sec:sec3}

The Fourier domain acceleration search method is implemented as a matched filtering process, where the Fourier response of a long time series is convolved in the Fourier domain with a number of short Finite Impulse Response (FIR) templates. Using the convolution theorem, convolution then becomes a complex multiplication, which is a local operation and is highly parallelisable. For pulsar searches, the resulting correlated response must be searched for candidates. This is done by calculating the power spectrum, and comparing the resulting Fourier power to a pre-calculated threshold. The power spectrum that results from the correlated output forms a 2-dimensional plane of frequency and frequency derivative. This plane is referred to as the \(f - \dot{f}\) plane. 

\begin{figure}[htb!]
\centering
\includegraphics[width=0.8\linewidth]{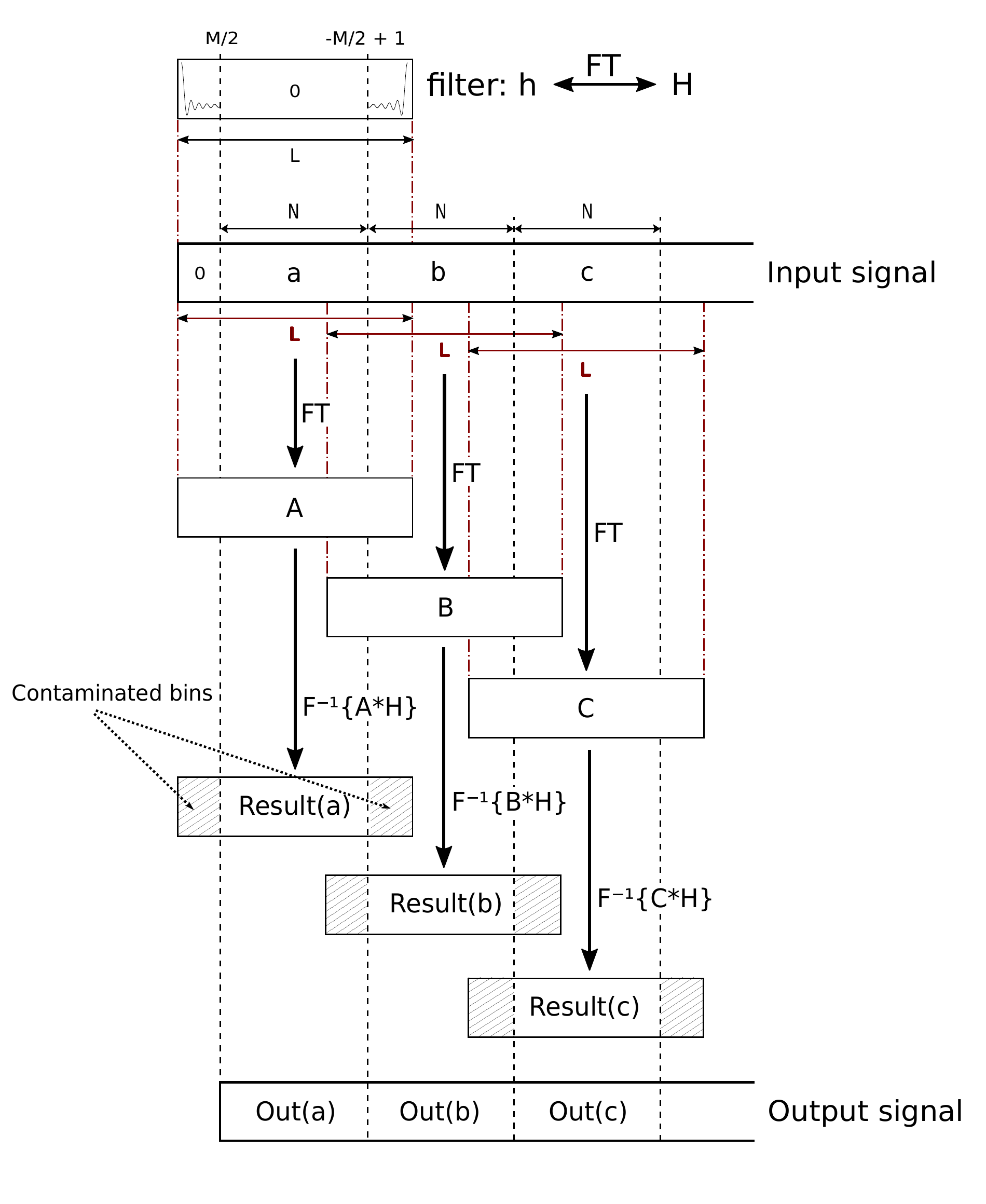}
\caption{The Overlap and Save method in FDAS. The complex filter h is placed at the edges of an array of length L, so as to be centered at bin 0, and its Fourier transform is represented by H. Blocks of length L (indicated by the red double arrows, and dash-dotted lines) are picked from the signal, overlapping continuous length-N segments (a, b, c, separated by dashed lines) by M/2 bins at each end. After Fourier transform (FT) of the blocks to form A, B, C, complex multiplication with H (indicated by *), and inverse Fourier transform, the edges in the results are discarded and the remaining blocks Out(a), Out(b), Out(c) concatenated to form the output signal.}\label{fig:f1}
\end{figure}

The Fourier response of the signal must be Fourier-transformed prior to the complex multiplication, and inverse transformed after. However, the convolution theorem expects that the input signal and template are of equal length, which is not generally the case. Furthermore, the FFT assumes that the signal is periodic with a period equal to the length  of the input, \(\boldsymbol{N}\), and the convolution performed via the FFT is cyclic. The cyclic nature of the convolution causes contamination of \(\boldsymbol{M}\) bins in the output signal, where \(\boldsymbol{M}\) is the width of the template. These issues can be treated by zero-padding of the input and template to the size of their linear convolution, \(\boldsymbol{N + M - 1} \). This would result in performing a long FFT for each template. To avoid this, an overlap and save method is used \citep{Press:1992:NRC:148286}, which works on small blocks of the signal independently, and accounts for the contamination in each block, as well as for the continuity of the convolution along the signal. A schematic representation of the process is shown in figure~\ref{fig:f1}.

As mentioned in section~\ref{sec:sec2}, the acceleration search templates are complex, and the phases need to be preserved to enable coherent recovery, so the filters are centered at bin zero. In order to perform the Fourier domain convolution, zeros are added in the middle of the \(L\)-length filter array between \(\boldsymbol{M/2}\) and \(\boldsymbol{L-M/2 + 1}\), where each half of the template is placed, as illustrated in the h filter block of figure~\ref{fig:f1}. After an initial zero-padding at the beginning of the input array, which is done to avoid the wrap-around pollution, equally sized blocks of the signal are picked for FFT convolution, each one starting from a region that overlaps the previous block by half the filter points. This ensures the continuity of the operation along the signal, and provides the necessary padding for the aliasing effect. After the convolutions are performed, the contaminated edges of each block can be discarded, and the blocks with the remaining useful output points are concatenated. An additional advantage of this method is that it allows one to chose the optimal block size for maximum performance of the FFT. 

\begin{figure}[htb!]
\centering
\includegraphics[width=\linewidth]{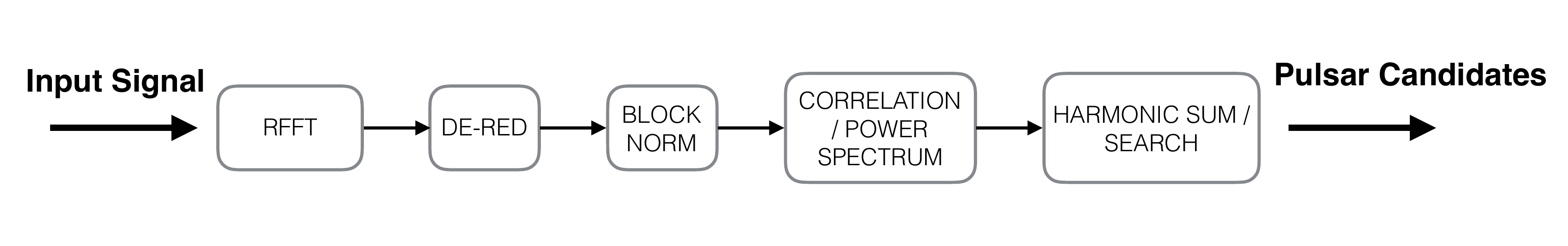}
\caption{The acceleration search process.}\label{fig:f2}
\end{figure}

Computationally, the acceleration search process is broken into a number of individual steps, shown in figure~\ref{fig:f2}. First, a real to complex FFT is performed on the input time series, and low frequency (red) noise is removed with the use of a local median filter \citep[see e.g][for a description]{2017MNRAS.467.1661V}. Next, the signal size is divided into blocks, and individual blocks are picked according to the overlap and save method. After a short step of normalization on the data of each block, the matched filtering is performed by applying a forward complex-to-complex FFT, complex multiplication with a number of pre-calculated templates, and inverse FFT of all the blocks that result from the block-filter pairs. The last step is to calculate the power spectrum, and search for peaks of Fourier power. Pulsar signals have the form of pulsed waveforms with an often very narrow duty cycle, which results in the distribution of their Fourier power across a number of harmonics. In order to increase the probability of detection, a harmonic summing step is typically performed before the search, by adding the power that is distributed in the harmonics  to that of the fundamental. The work described in this document deals only with the correlations and power calculations and does not include the steps of normalization, harmonic summing and candidate selection. 

In order to compensate for the loss of sensitivity that occurs for frequencies that fall outside Fourier bin centres (known as `scalloping loss'), a method known as `interbinning' is commonly used in the Fourier analysis of pulsar observations \citep{1993ispu.conf..372M,2004hpa..book.....L}. This involves the computation of the Fourier response at half-bin frequencies, which is approximated from the interpolation of the amplitudes of the two neighbouring bins, and given by

\begin{equation}\label{eq:f-interp}
\mathcal{F}_{k+\frac{1}{2}} \simeq \frac{\pi}{4} (\mathcal{F}_k - \mathcal{F}_{k+1}).
\end{equation}

\noindent Interbinning reduces the maximum loss in Fourier amplitude to \(\sim \) 7\%, from a potential 36\% loss, for a relatively low computational cost, and for this reason it is usually incorporated in periodicity searches. For the Fourier domain acceleration search, a 2 bin Fourier interpolation \citep{1538-3881-124-3-1788} can be used by doubling the template and signal resolution using interleaving zeros prior to the correlations, or  interbinning can be explicitly implemented, by applying equation \eqref{eq:f-interp} to the correlated fourier amplitudes. The latter has the advantage of performing the Fourier domain computations on the initial signal resolution, while the former is more accurate, but operates on double the number of points throughout the process. Our GPU application uses the interbinning technique on the Fourier amplitudes after the correlations. 

\section{GPU Implementation} \label{sec:sec4}

We have developed and tested two GPU implementations for the FDAS algorithm. The first, which we use as a reference, makes use of the cuFFT library for the Fourier operations, along with custom functions that perform the convolution and power spectrum operations. The second implementation incorporates a short length custom FFT that helps perform all the correlation and power spectrum processes using fast on-chip resources. The details of each implementation are described in this section. Our application is targeted to time domain processing for radio telescopes, and applies to a limited input size, that does not exceed the GPU on-board memory resources. Current NVIDIA technology provides enough memory to enable the processing of signals with up to \(2^{24}\) input samples and 300 correlation templates or close, which is within  the specifications of the SKA for acceleration searches.

\subsection{Using cuFFT}
\begin{figure}[htb!]
\centering
\includegraphics[width=0.8\textwidth]{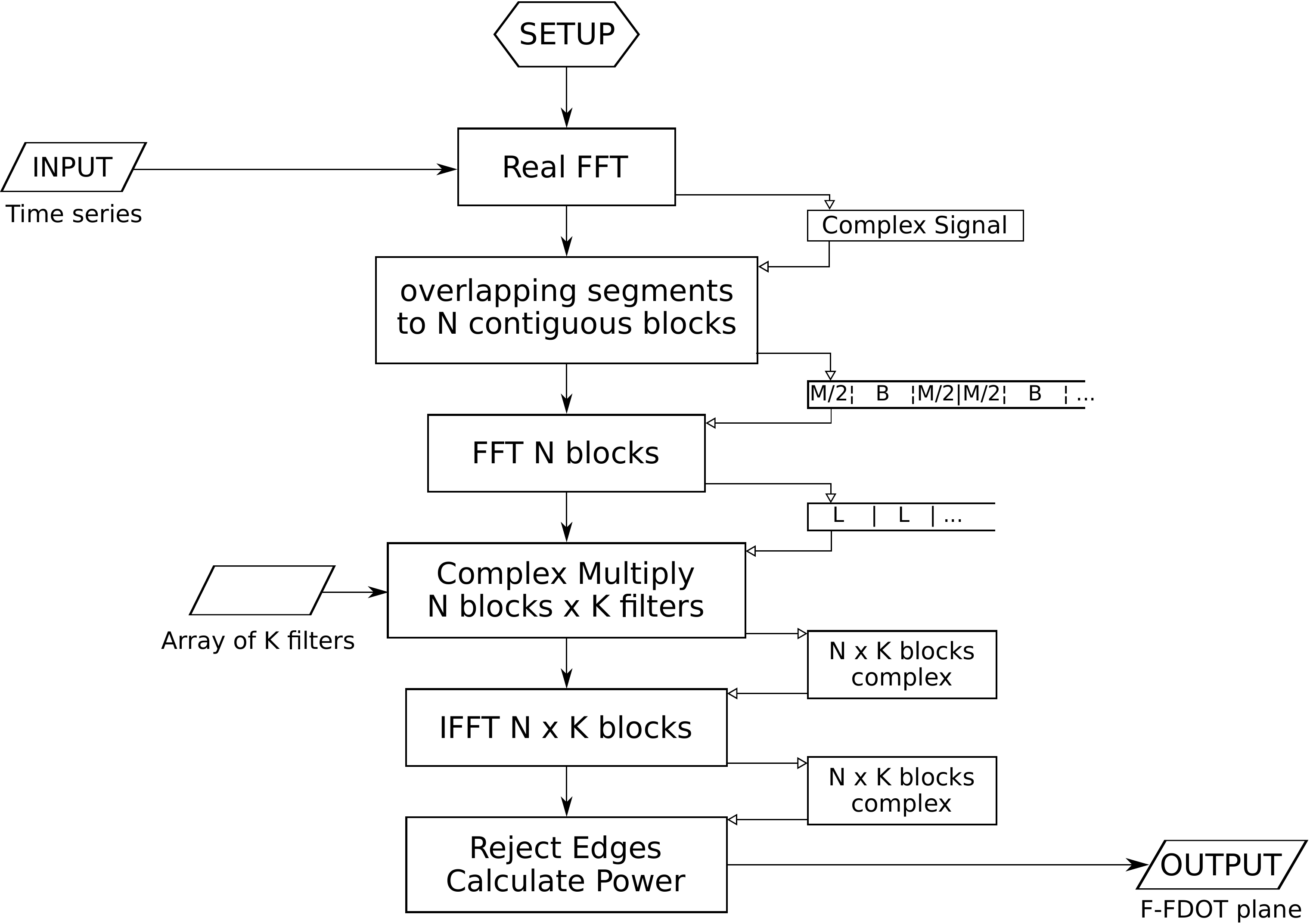}
\caption{Workflow of Fourier Domain matched filtering on GPU using the cuFFT library. After an initial setup, the Real FFT is applied on the input series, producing the complex signal array. Segments overlapping by half the filter length (M/2) are picked from this array according to the overlap and save method, and placed contiguously into a separate array as N blocks of length L = B + M, where M is the length of our template, and B is the number of useful output points to recover from each block. An array of K filters stored in memory is loaded to perform element-wise complex multiplication on a total of N\(\times\)K L-length blocks, with the result written to a N\(\times\)K\(\times\)L complex array. An inverse FFT is performed on this array, and finally the edges rejected and the Fourier powers are calculated resulting in the real f-fdot plane, an array of total length N\(\times\)K\(\times\)B.} 
\label{fig:f3}
\end{figure}

At the core of the matched filtering process described in section~\ref{sec:sec3} are a set of forward and inverse Fourier transforms. As a first step, we used the NVIDIA cuFFT library, which provides a highly optimised implementation of FFT routines written in CUDA for NVIDIA GPUs, with an easy to use host interface. The rest of the steps were also performed on the GPU using CUDA device code. The GPU workflow is shown in the schematic diagram of figure \ref{fig:f3}.

The input signal that represents a time series is a one dimensional array of \(n\) 32-bit floating point elements. This array is transformed to the Fourier domain using a Real-to-Complex cuFFT routine. Overlap and save is implemented by way of copying \(N\) overlapping segments from the Fourier domain input to a new array in contiguous positions. The segments are of length \(L = B + M \), where \(B\) is the size of the useful output points from each segment and \(M\) is the length of the filter. The remaining computations can then take place in independent memory regions. Each segment is transformed with a Forward Complex-to-Complex FFT using a batched cuFFT routine, where multiple segments are transformed on the GPU with one call in parallel. Each transformed segment then is element-wise multiplied with each of \(\boldsymbol{K} \) acceleration templates, producing an \(f-\dot{f}\) of complex numbers that includes the contaminated regions. Next, the segments are inverse Fourier transformed with a batched routine, this time the number of batches is multiplied by \(\boldsymbol{K}\). Finally, the result is processed by a device kernel that computes the Fourier power for the whole   \(f-\dot{f}\) plane and stores the useful \(B\) elements of every segment to a floating point array contiguously.

\begin{figure}[htb!]
\centering
\includegraphics[width=0.7\textwidth]{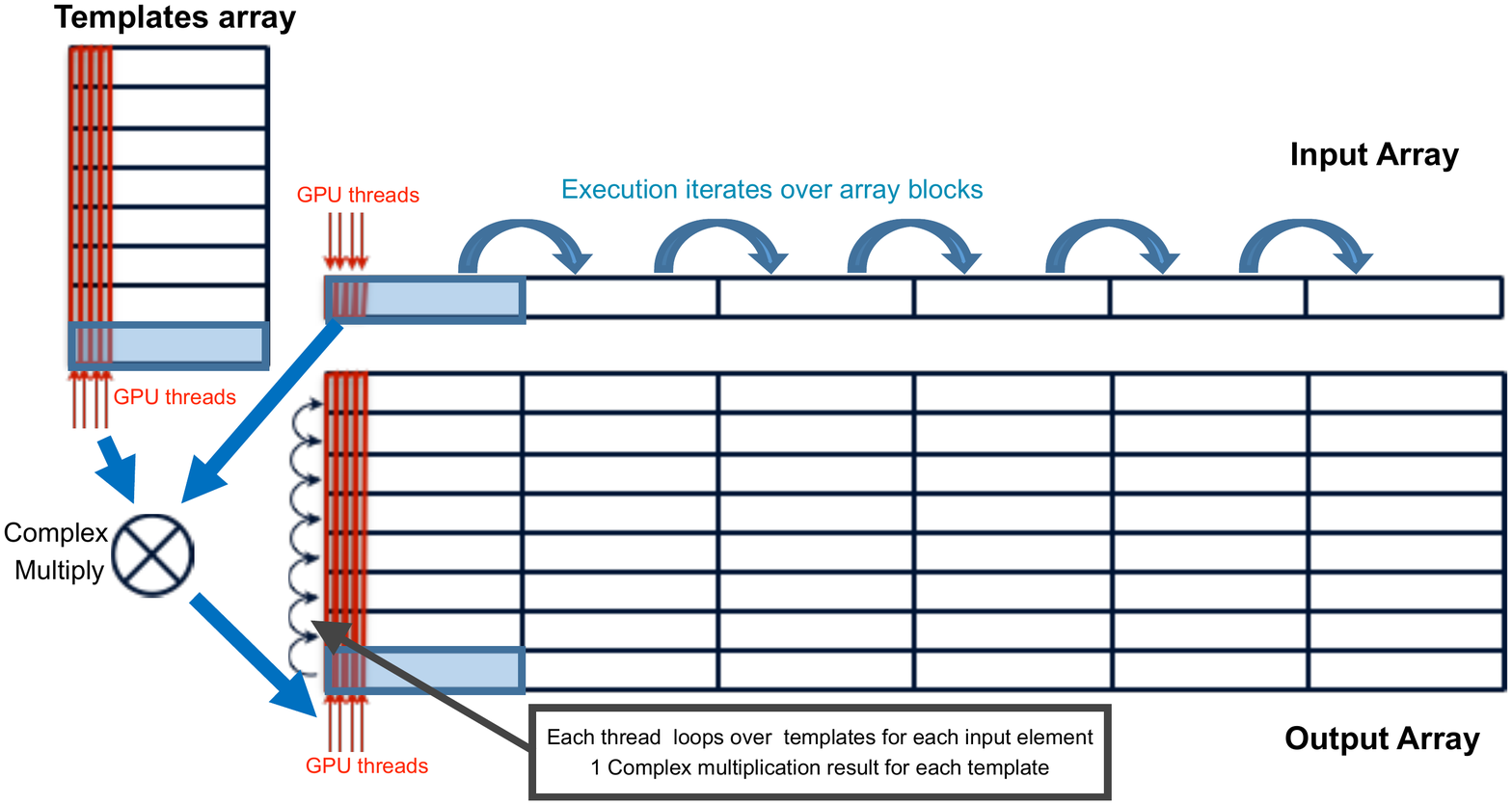}
\caption{Schematic representation of complex multiplication GPU kernel using arrays of registers and iteration over templates.} 
\label{fig:f4}
\end{figure}

The complex multiplication GPU kernel is structured in two ways. In the first scheme (figure \ref{fig:f4}) a one dimensional GPU  grid of threads the size of a single segment, loads all templates, with each thread loading a `column' of the stacked templates on device registers.  It then loops along the signal and across all templates to perform the complex multiplications. For each signal segment, the algorithm loads the input and then each thread iterates over its array of template elements, performing one complex multiplication per element, and writing the result back to the device global memory. The process is repeated for every signal segment. This scheme allows for maximum memory reuse with on-chip resources, but suffers from register and shared memory pressure, meaning that each thread block consumes a significant amount of GPU multiprocessor resources, that limit the amount of blocks that can be run simultaneously on the device. There can also be register spillage to the global memory, that induces a significant overhead to the execution. Optimisations were done for this scheme to avoid such spillage, as well as to reduce memory dependencies using loop unrolling and instruction level parallelism techniques. To increase the memory read speed, we made use of the read-only data cache which is available with NVIDIA devices of the Kepler generation and above, and is known as the texture cache for previous generations. Data alignment to 32 bit floats was also used with shared memory transactions to improve bandwidth.  

\begin{figure}[htb!]
\centering
\includegraphics[width=0.7\textwidth]{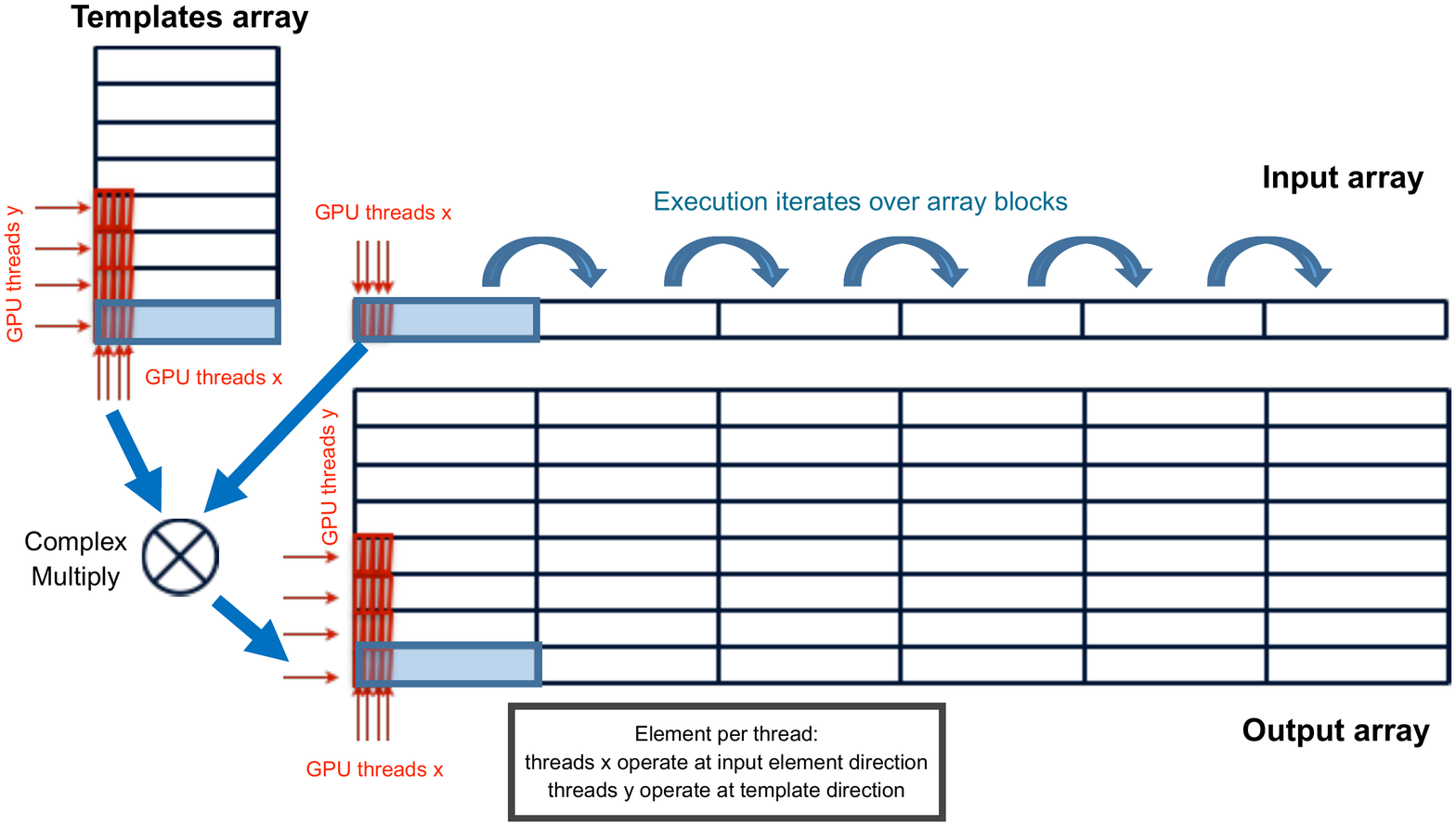}
\caption{Schematic representation of complex multiplication GPU kernel using a two-dimensional GPU compute grid.} 
\label{fig:f5}
\end{figure}

The second scheme is shown in figure \ref{fig:f5}. Instead of looping through the templates, this scheme utilises a two-dimensional grid where the vertical direction of the grid represents the template number. The grid then operates on a thread-per- \(f-\dot{f}\) element basis. This scheme has very high multiprocessor occupancy, by utilising a very high percentage of the computing cores in parallel. 

Both schemes are memory bandwidth limited, and have similar execution speeds in general, but the first shows relative gains over the second when a GPU architecture with higher number of available registers is used. Because of the use of the cuFFT library and the low arithmetic intensity in this kernel, the memory bandwidth limitation cannot be improved significantly, but the first scheme was preferred because it proved to have more potential for improvement in performance with increases in device on-chip resources. This step, using currently available hardware, consumes approximately 20\% of the total execution time.

The final step  is to compute the power spectrum of the complex \(f-\dot{f}\) plane. This is also a bandwidth limited operation which only involves two multiplications and one addition per element read / write. It accounts for about 35\% of the execution time.

This implementation of FDAS is dominated by the cuFFT operations, that cost around 42\% of the execution time, but is also limited by the memory bandwidth required to read and write data at intermediate stages, that is imposed by the use of cuFFT. Under these conditions the template length has been chosen to produce the best cuFFT performance, and other GPU parameters, such as grid and thread block size for GPU kernel execution, as well as the kernel structure, were tuned to produce optimal overall performance, which also aligns with the cuFFT requirements. 

\subsection{Using a custom FFT on the GPU shared memory}
To overcome the  limitations of the first implementation, we have developed a custom GPU FFT code and incorporated it into the algorithm. In order to avoid transferring data to and from GPU global memory, we aimed at performing the FFTs for a full segment entirely on the chip's shared memory, so that the result can be immediately used in the other steps, which are performed within the same kernel. For this reason, the template size in the current implementation is limited to 1024 points. The matched filtering algorithm was applied using the first complex multiplication kernel scheme, but it was found that there was no benefit in pre-storing the templates. A complex conjugate symmetry that has been observed previously between the positive and negative acceleration templates, was exploited so that half of the templates need not be transferred from the global memory, but can be applied on-the-fly as the complex conjugate of their symmetric template contained in the other half. The power calculations are also done on shared memory data, and the clean, concatenated result is stored in the global memory at the end. The grid of thread blocks with this method extends to the length of the Fourier domain input, and the number of threads in a block is equal to the FFT size, hence there is no need to loop over the signal segments.  The remainder of this section describes the FFT algorithms that were investigated for this implementation.

The discrete Fourier transformation (DFT) of a signal $x$ is given by 
	\begin{equation}
	\label{eqa:DFT}
		X_m=\sum_{n=0}^{N-1} x_n \textup{e}^{-i2\pi nm/N}=\sum_{n=0}^{N-1} x_n\, \omega_{N}^{nm}\,,
	\end{equation}
	
	where $X_m$ is a signal in the \textit{frequency-domain}, $x_n$ is a signal in the \textit{time-domain},  $N$ is the signal length or DFT length and the exponential factors $\omega_{N}^{nm} = \textup{e}^{-\textup{i}2\pi nm/N}$ are called \textit{twiddle} factors.
	The inverse discrete Fourier transformation (IDFT) is 
	\begin{equation}
	\label{eqa:DFT}
		x_m=\sum_{n=0}^{N-1} X_n \textup{e}^{i2\pi nm/N}=\sum_{n=0}^{N-1} X_n\, \omega_{N}^{-nm}\,.
	\end{equation}
	
	We can also express the discrete Fourier transformation as a matrix multiplication
	\begin{equation}
	\label{eqa:matrixDFT}
	X = \bb{F}x \, ,
	\end{equation}
	where $\bb{F}$ is a matrix of twiddle factors called the Fourier matrix $\bb{F}\in\mathbb{C}^{N \times N}$, $X$ and $x$ are vectors $X,x\in\mathbb{C}^{N}$ in the frequency and time domain respectively. 
		
The discrete Fourier transformation is an operation of complexity $O(N^2)$. The Fast Fourier Transform or FFT allows us to calculate the DFT with a lower number of operations and higher precision. The original FFT algorithm was published by Cooley \& Tukey \citep{Coo-Tuk:1965:FFT}, and since then, many different FFT variants have been published \citep{Loan:1992:FFT_Computational_Frameworks}. 

The Fourier matrix in equation \eqref{eqa:matrixDFT} could be expressed as a multiplication of factorization matrices. In the case of the Cooley-Tukey algorithm, the Fourier matrix $\bb{F}_N$ can be written as 
	\begin{equation}
	\label{eqa:CTfactorization}
	\bb{F}_{N}=\bb{A}_{t}\bb{A}_{t-1}\ldots\bb{A}_{1}\bb{P}_N\,,
	\end{equation}
where $\bb{A}_q$ are factorization matrices, $\bb{P}_N$ is a permutation matrix responsible for the re-ordering of time-domain vector $x$. The index $t$ is related to the DFT length $N$ and if we restrict ourselves to radix-2 algorithms it is given by $N=2^t$. The factorization matrices $\bb{A}_q$ depend on the chosen FFT algorithm as well as on the presence of the perturbation matrix $\bb{P}_N$\footnote{autosort algorithms like Stockham algorithm do not require a perturbations matrix}. For the exact definition of the factorization matrices and more details about matrix representation of FFT algorithms we refer the reader to  \citep{Loan:1992:FFT_Computational_Frameworks}. 
	
One more distinction has to be made with FFT algorithms. Each FFT algorithm has two possible variants, namely decimation in time (DIT) and decimation in frequency (DIF). These variants reflect the manner in which we divide input data. The relevance of these two variants for convolution lies in the position of the perturbation matrix $\bb{P}_N$. The expression for the DIT FFT algorithm is given in equation \eqref{eqa:CTfactorization}. To obtain the DIF we only need to transpose equation \eqref{eqa:CTfactorization}. This will give us
	\begin{align}
	\bb{F}_{N}&=\bb{A}_{t}\bb{A}_{t-1}\ldots\bb{A}_{1}\bb{P}_N\,,&\\
	\bb{F}^T_{N}=\bb{F}_{N}&=\bb{P}_N\bb{A}^T_{1}\ldots\bb{A}^T_{t-1}\bb{A}^T_{t}\,.&
	\end{align}

\noindent Using these in equation \eqref{eqa:matrixDFT} results in
	\begin{align}
	\label{eqa:FFTDIT}
	X&=\bar{\bb{A}}\bb{P}_{N}x\,,&\\
	\label{eqa:FFTDIF}
	X&=\bb{P}_N\bar{\bb{A}}^{T}x\,,&
	\end{align}
where we have replaced the factorization matrices with $\bar{\bb{A}}$. We can see that in the case of the decimation in time algorithm (eq. \eqref{eqa:FFTDIT}), the FFT algorithm represented by $\bar{\bb{A}}$ requires the time-domain vector $x$ to be reordered by the permutation matrix $\bb{P}_{N}$ and produces the correctly ordered frequency-domain vector $X$. The decimation in frequency algorithm on the other hand requires the in-order time-domain vector $x$ and produces vector $X$ which needs to be re-ordered by permutation matrix. The same holds for the inverse DFT. Since the convolution in the frequency domain is represented by point-wise multiplication, the result does not depend on the order of the elements within vectors which are being convolved, as long as both vectors are ordered in the same way. Thus we can eliminate the permutation matrix $\bb{P}_N$ from our FFT implementation by choosing the DIF FFT variant for the forward FFT and the DIT variant for the inverse FFT algorithm. This enables us to save memory transactions and decrease execution time.
	
We have examined and implemented three radix-2 FFT algorithms. These are: the Cooley-Tukey algorithm \citep{Coo-Tuk:1965:FFT}; the Pease algorithm \citep{Pease:1968:AFF:321450.321457}; and the Stockham algorithm \citep{Stockham:1967:FFT}. We have considered DFTs with lengths of power-of-two only. To perform the DFT we use two different algorithms. For the forward DFT we use the DIF variant of the Pease algorithm and for the inverse DFT we use the DIT variant of the Cooley-Tukey algorithm. 
	
For the calculation of the twiddle factors we use CUDA fast-math intrinsics\footnote{more in appendix D of NVIDIA CUDA programming guide \url{http://docs.nvidia.com/cuda/cuda-c-programming-guide}}. We reuse the twiddle factors by calculating two elements of the DFT transformation per thread. In the case of the inverse FFT we are also performing two transformations simultaneously, when exploiting the symmetry of the templates as mentioned earlier in this section, and this further increases the reuse of twiddle factors, which are calculated once.

The Cooley-Tukey FFT algorithm is perhaps the best known FFT algorithm. It requires re-ordering in order to produce the correct result. The memory access pattern for the algorithm is given by a butterfly diagram. During this operation, results of smaller DFTs are combined into larger DFTs. For smaller butterflies this induces shared memory bank conflicts. However, by using thread shuffle instructions we can remove these bank conflicts and also reduce synchronization requirements between threads.
	
We have also implemented the Pease FFT algorithm in the DIF variant which requires re-ordering of the output (frequency-domain) vector. The Pease algorithm has the interesting feature that its data access pattern remains constant across all iterations. For the radix-2 variant this memory access pattern is favourable for a shared memory implementation of the algorithm.
	
Our final arrangement used for convolutions is a combination of the Pease DIF FFT algorithm for forward FFT transformation and the Cooley-Tukey DIT FFT algorithm for inverse transformation. A combination of the DIF and DIT algorithm allowed us to eliminate the re-ordering step, which is otherwise necessary to produce correct results, and thus to reduce the number of operations performed and the execution time. When using the FFT without the re-ordering step we have to use the same algorithm also for the templates transformations. We have used the Pease DIF FFT algorithm instead of the Cooley-Tukey DIF FFT algorithm, because the Cooley-Tukey algorithm had higher register usage.
	
Using the shared memory FFTs in the matched filtering step eliminated the device memory dependence of the algorithm, and the execution performance is now limited by the efficiency of the use of on-chip resources and by synchronisation.

\section{Experimental Results} \label{sec:sec5}
We have run a series of tests on our custom code, in order to establish its signal recovery capability, as well as to assess computational performance.  As described in section~\ref{sec:sec3}, the full acceleration search process in PRESTO consists of several functional parts, which are shown in figure~\ref{fig:f2}. The FDAS GPU implementation optimises only the matched filtering part, and this is the subject of the comparisons to follow. 

\subsection{Computational Performance}
We compared computational speed between PRESTO, PRESTO-GPU, and our FDAS GPU implementation with cuFFT and with our custom FFT for varying input signal sizes, and a varying number of correlation templates. Only the real FFT, correlations and power spectrum calculations were timed in all cases. The tests were performed both with and without use of interbinning. PRESTO was run using 20 OpenMP threads on the 10 CPU cores (20 Hyperthreading cores) of an Intel Xeon E5-2650. Our GPU codes were run on the NVIDIA Maxwell architecture M40 card, as well as on the dual-GPU NVIDIA K80 (Kepler architecture) and M60 (Maxwell) cards using only a single of the two GPUs on-board (when using each GPU of the dual cards to process a different time series, the execution speed of the total number of independent time series is doubled).  Preliminary measurements on the latest NVIDIA architecture, Pascal, with the Tesla P100 card are also presented. 

We first compare our basic GPU implementation to PRESTO and PRESTO-GPU. We then examine the performance of our custom GPU code against our basic GPU version, and finally compare this to PRESTO and PRESTO-GPU. The comparison with PRESTO-GPU is only meant to give an indication of the difference in execution time, and for this reason we only used the M40 card which is a representative high-end single-GPU card of the Maxwell architecture. Table \ref{table:t1} lists the main specifications of the GPU hardware.
\begin{table}[htbp]
\caption{Hardware specifications}
\centering
\begin{tabular}{lcccc}
\hline
  & K80 & M60 & M40 & P100 \\ \hline
\# of GPUs on card\footnote{Where there are 2 GPUs, only 1 is used in the experiments.} & 2 & 2 & 1 & 1 \\ 
\# of Streaming Multiprocessors (SMs)  & \(2\times 13 \) & \(2\times 16 \) & 24 & 56 \\ 
Total \# of cores  & \(2\times  2496\) & \(2\times 2048 \) & 3072 & 3584 \\ 
Base / Max Core Clock (MHz) & 560 / 875 & 899 / 1178 & 948 / 1114 & 1189 / 1328\\ 
Main memory bandwidth (GB/s) & $2\times 240$  & $2\times 160$ & 288 & 732 \\ 
Main memory size (GB) & $2\times 12 $ & $2\times 8$ & 12 & 16 \\
L2 cache size (MiB)& 1.5 & 2.0 & 3.0 & 4 \\ 
Peak Shared memory bandwidth (est.)\footnote{The shared memory bandwidth was estimated using the formula: \newline
	 \indent BW (bytes/s) =  (bank bandwidth (bytes)) $\times$ (clock frequency (Hz)) $\times$ (32 banks) $\times$ (\# multiprocessors). } (GB/s) & $2\times 2712$ & $2\times 4494$ & 6374 & 9519 \\ \hline
\end{tabular}
\label{table:t1}
\end{table}%

It should be noted that, because the cuFFT requires \(1.5 \times \) the amount of GPU global memory storage than the custom code, due to the need to store the result of the inverse complex FFTs, some trials with signal size of \(2^{23}\) and a high number of templates could not be executed with this version, because they exceeded the available GPU memory capacity. These can be seen as missing parts of lines in the results figures. This is a limitation of the cuFFT version. The custom code allows larger \(f - \dot{f} \) plane sizes to be processed. As the GPU architecture evolves, it is expected that memory storage will not be a limiting factor in future generations for our problem sizes.

\subsubsection{Comparisons of FDAS-cuFFT to PRESTO} \label{sssec:sssec61}
In this section we report on the comparison in computational performance of our FDAS with cuFFT code against both the CPU OpenMP and GPU version of PRESTO.

PRESTO was run using 20 OpenMP threads on  a 10-core Intel Xeon E5-2650 (Haswell), and only the correlations were timed (including the long real FFT), for input signal sizes of \(2^{21}, 2^{22}\), and \(2^{23}\) (the closest powers of 2 for 2, 4, and 8 million samples), with a varying number of templates up to 256. Presto GPU and FDAS cuFFT  run the same test on an NVIDIA M40 GPU card. 

\begin{figure}[htb!]
\gridline{\fig{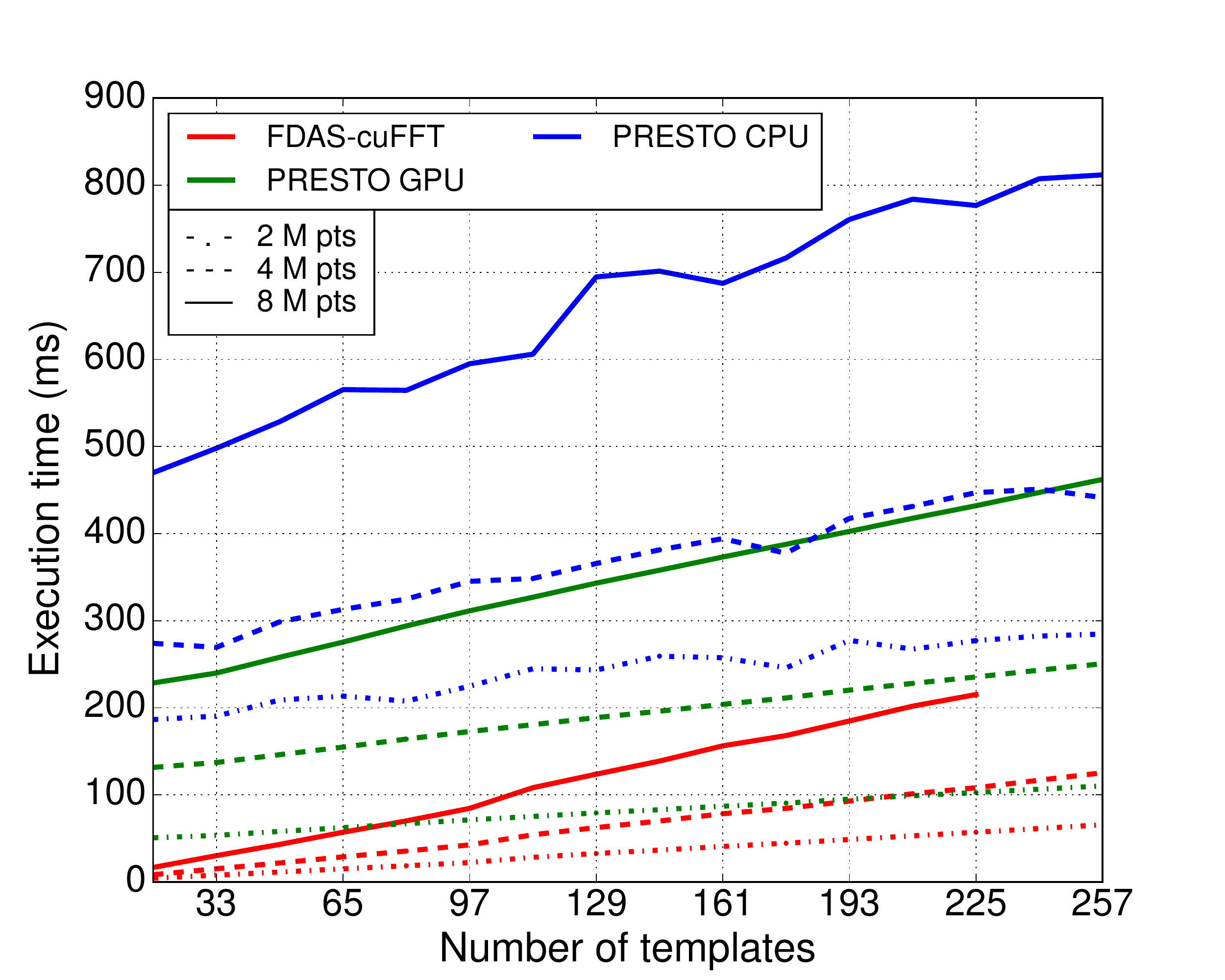}{0.5\textwidth}{(a)} \label{fig:f6a}
\fig{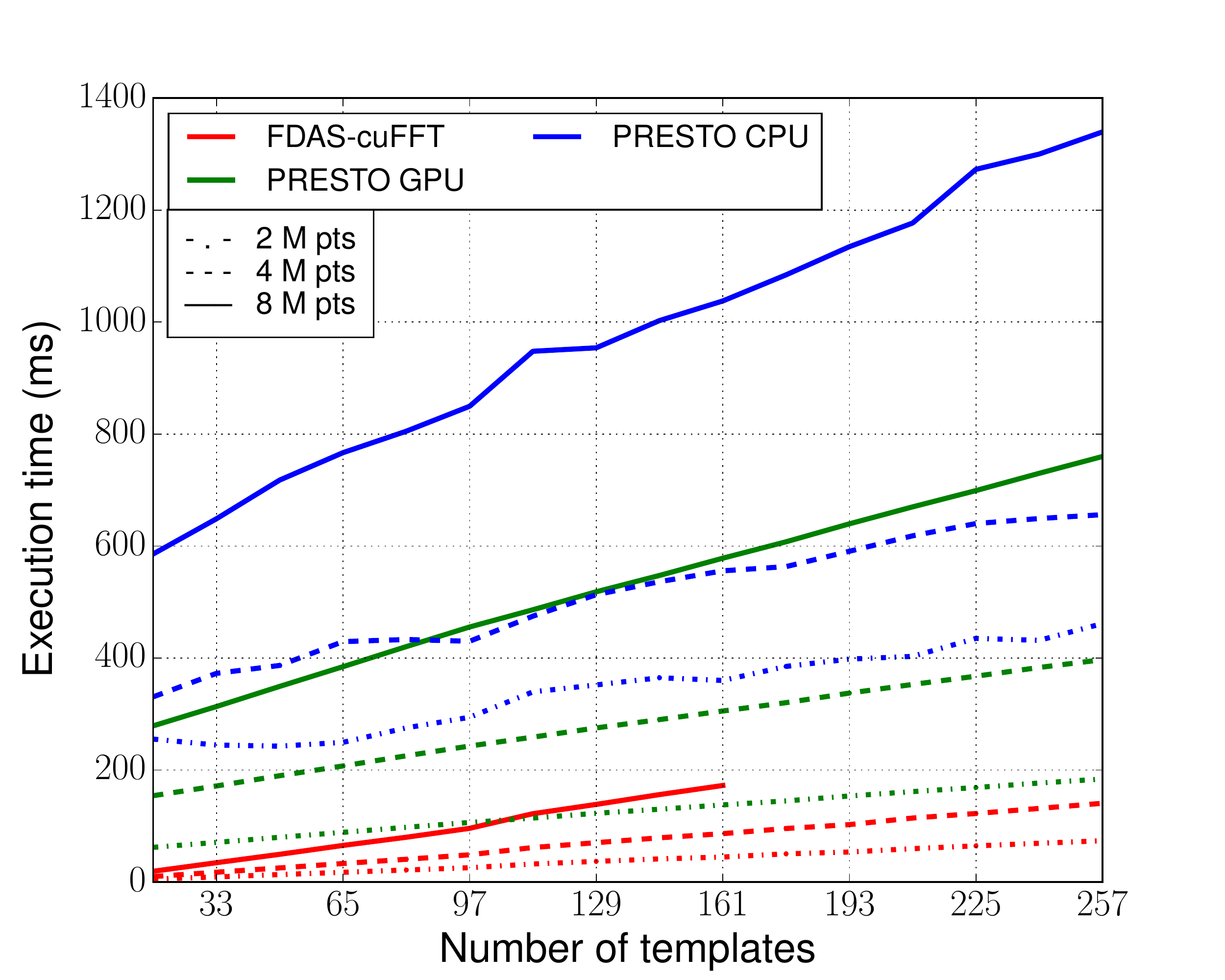}{0.5\textwidth}{(b)} \label{fig:f6b}
}
\gridline{
\fig{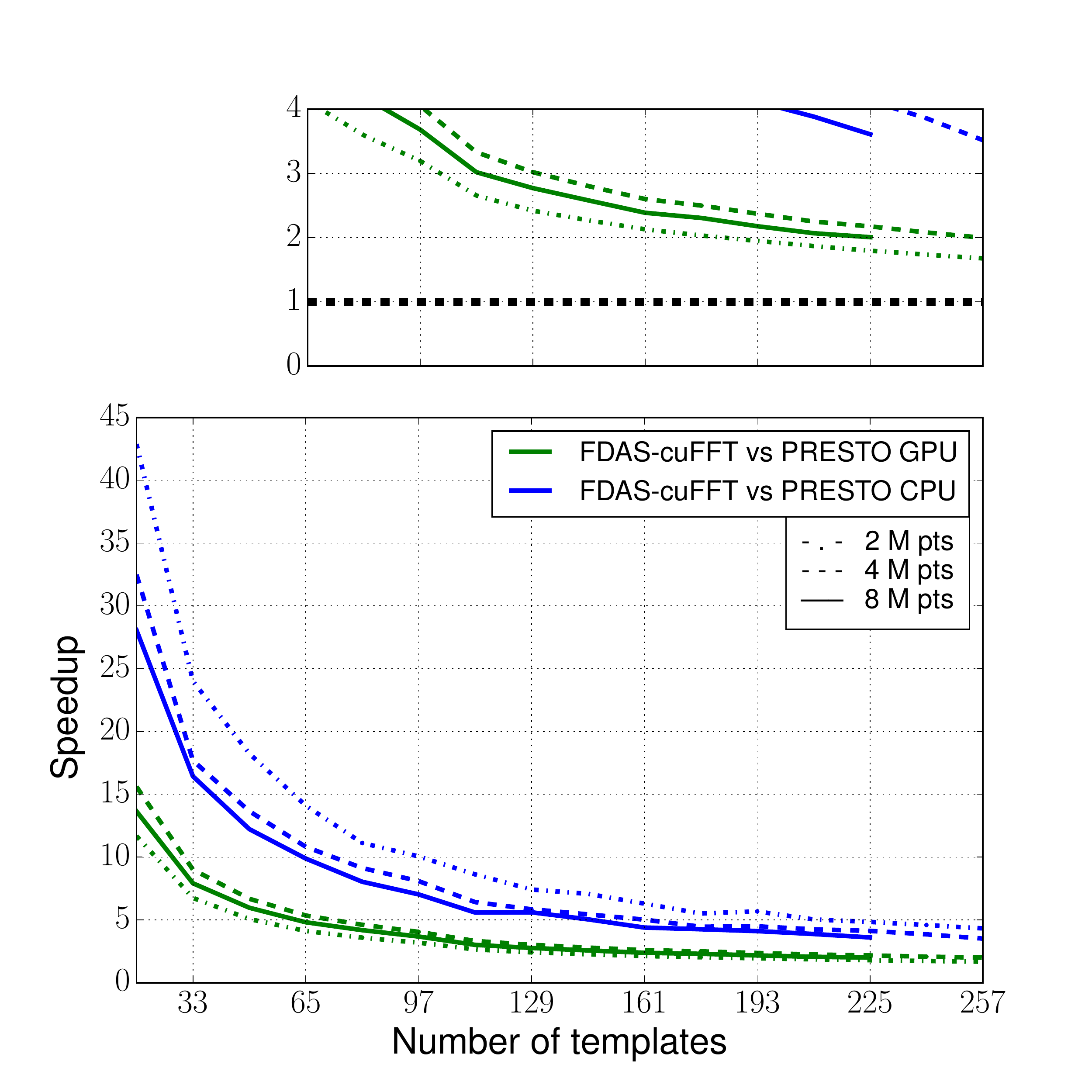}{0.5\textwidth}{(c)} \label{fig:f6c}
\fig{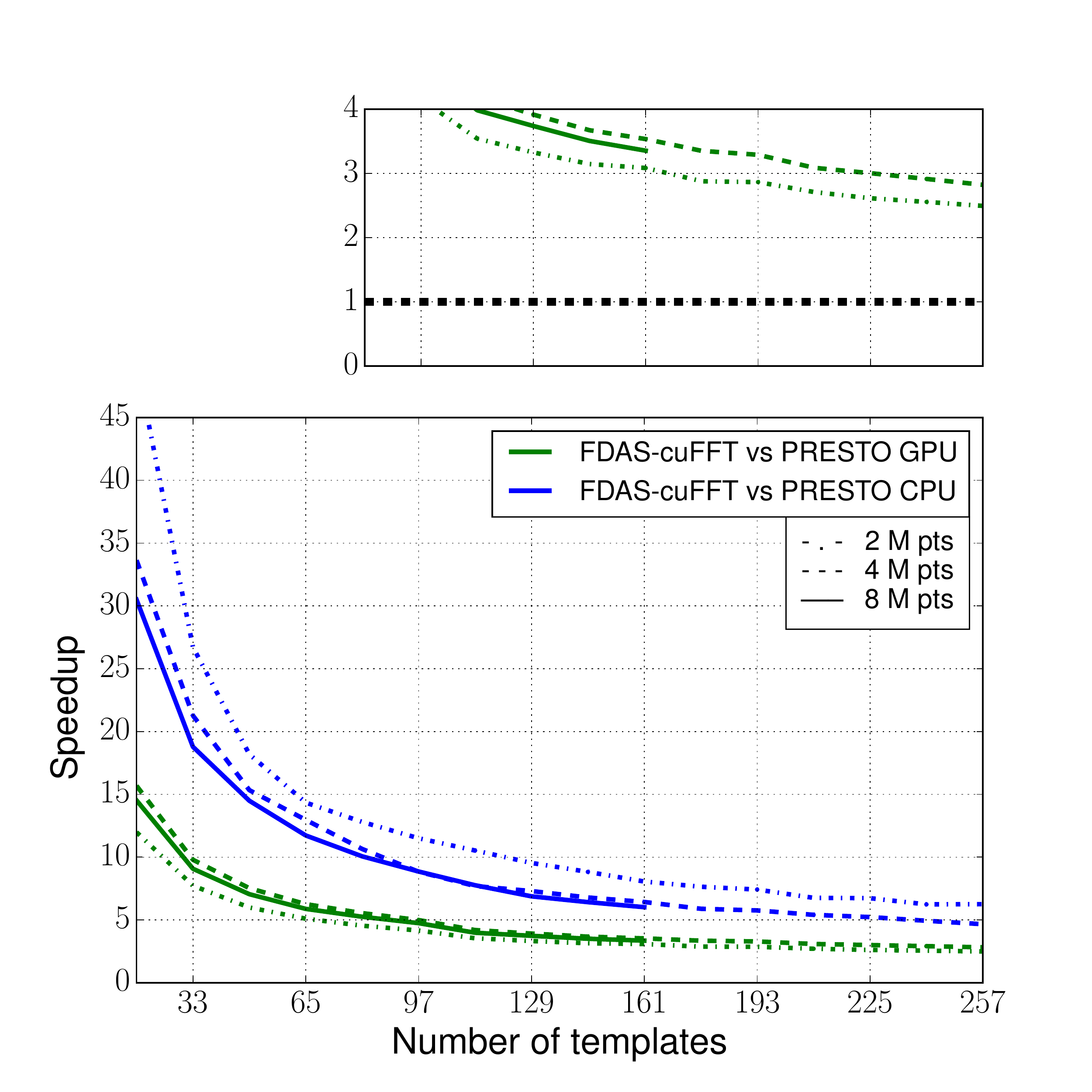}{0.5\textwidth}{(d)} \label{fig:f6d}
}

\caption{Comparison of FDAS-cuFFT to the PRESTO CPU and GPU versions on the NVIDIA M40. (a) and (b): Measured execution time with increasing number of filter templates, (c) and (d): Speedup of FDAS-cuFFT compared to PRESTO CPU and GPU, (c), (d) top: zoomed y-axis to show lower speedups in more detail, with black dashed horizontal line at the point where speedup = 1. The graphs on the left are for runs  without interbinning, and on the right with interbinning.}
\label{fig:f6}
\end{figure}

Figure~\ref{fig:f6} shows the execution time of each code along number of filter templates, and the gain in speed of  FDAS cuFFT from each of the PRESTO codes. The execution time of all three codes appears to be approximately linear with the number of templates, as a result of the similar blocked algorithm employed, which iterates over FFT blocks. This imposes approximately the same number of memory transfers, but on different memory subsystems. The method is inherently memory bandwidth dependent, therefore the difference in execution time relies heavily on memory bandwidth. The CPU version appears to have some significant variations from linearity at points, which are mainly the effect of the OpenMP parallelisation across filters. The speedup graphs show a characteristic drop in speedup. The steepness of the decrease is due to an approximately constant difference in execution time of the long real-to-complex FFT, which is run on the CPU in both PRESTO codes and is at the order of 30, 75, and 220 \(m s\)  for the signal lengths of  \(2^{21}, 2^{22}\), and \(2^{23}\) respectively. 

\begin{figure}[htb!]
\gridline{\fig{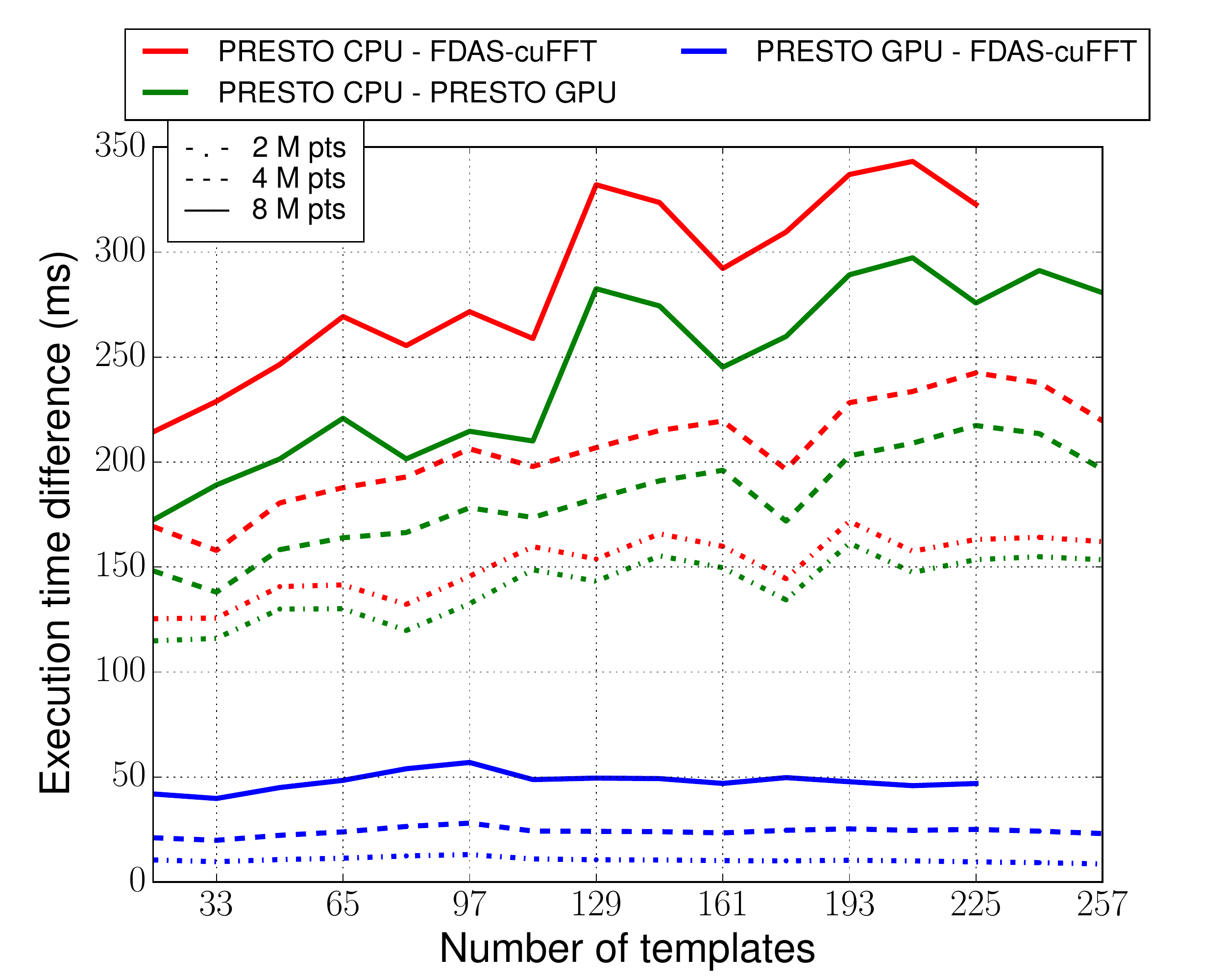}{0.5\textwidth}{(a)} \label{fig:f7a}
\fig{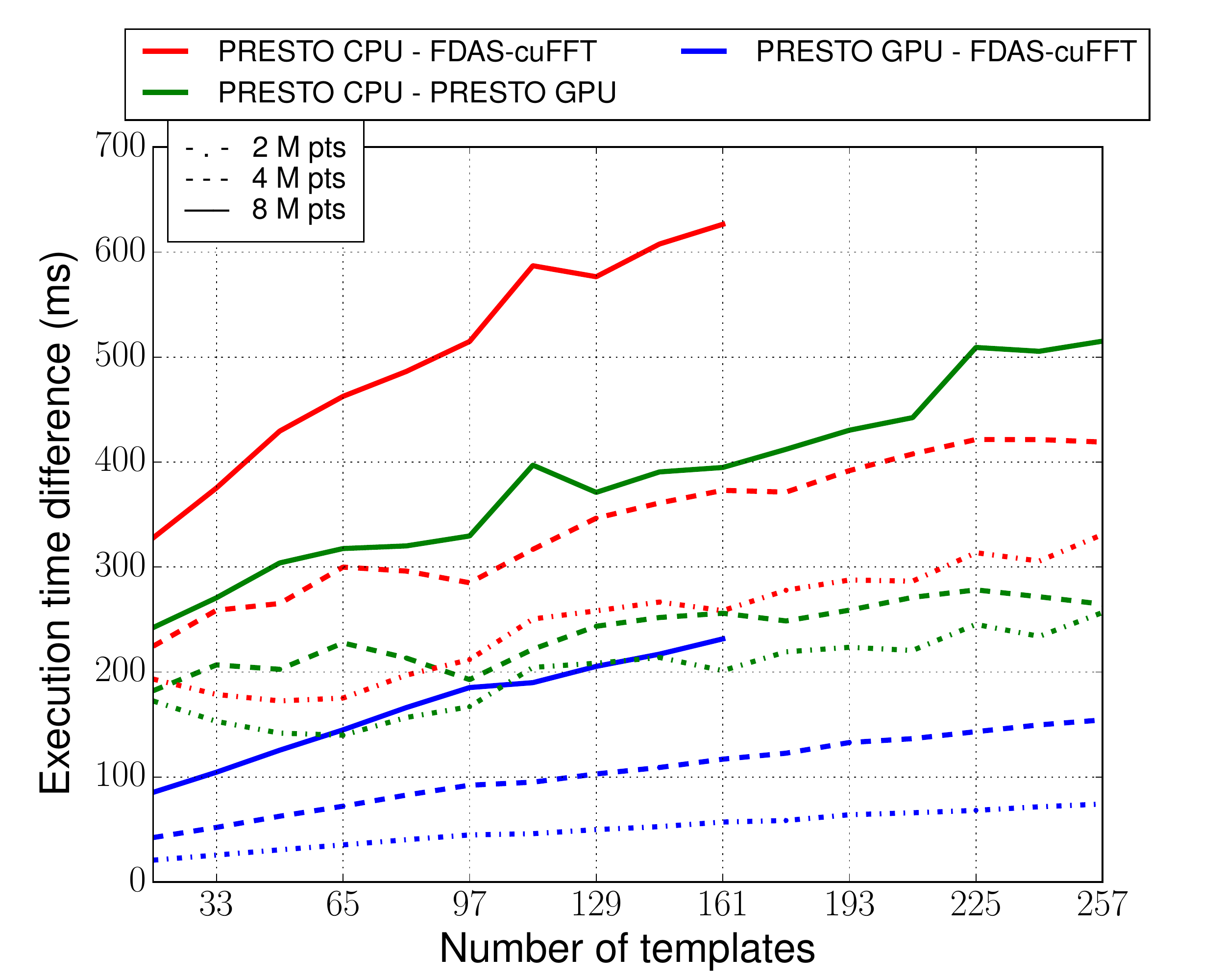}{0.5\textwidth}{(b)} \label{fig:f7b}
}
\gridline{\fig{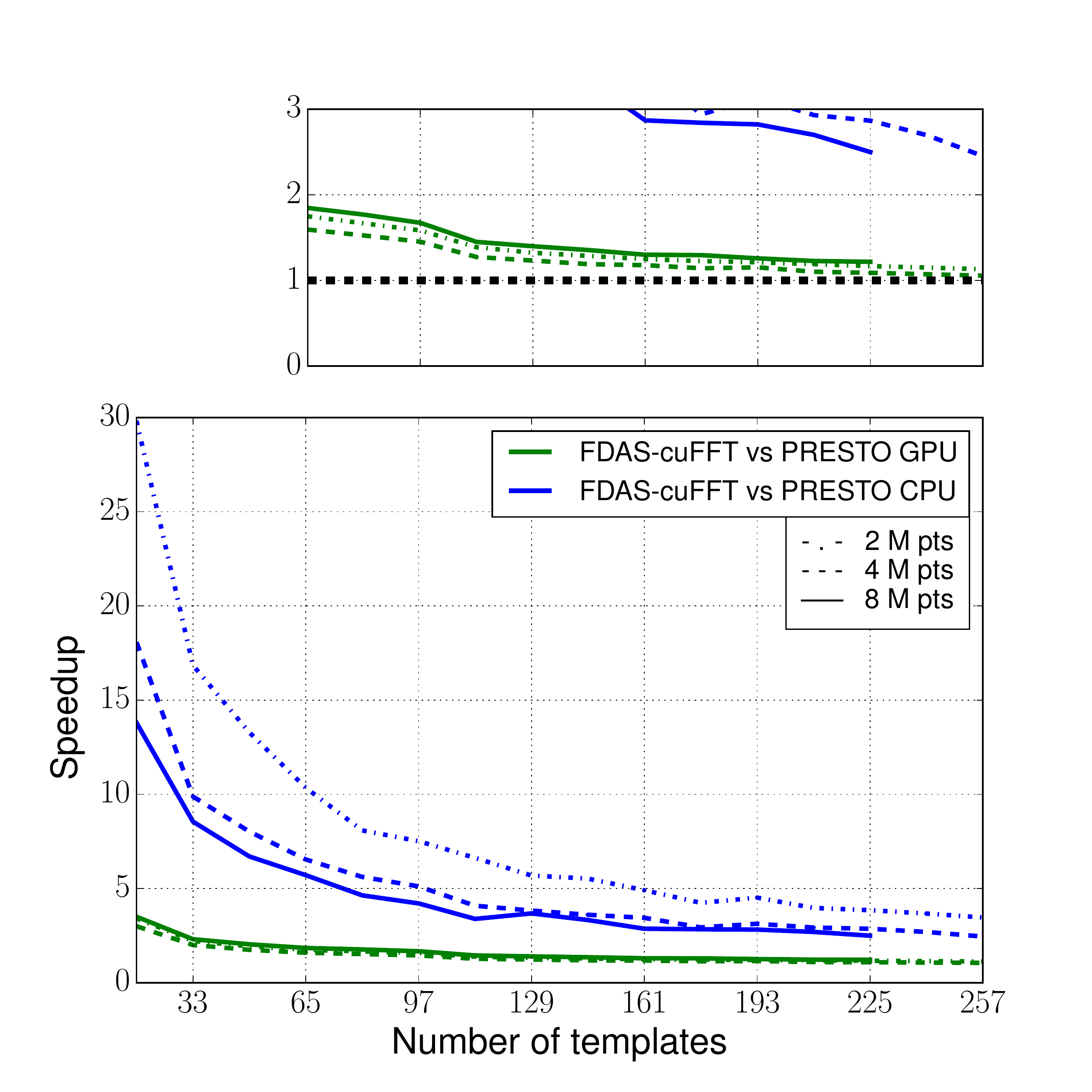}{0.5\textwidth}{(c)} \label{fig:f7c}
\fig{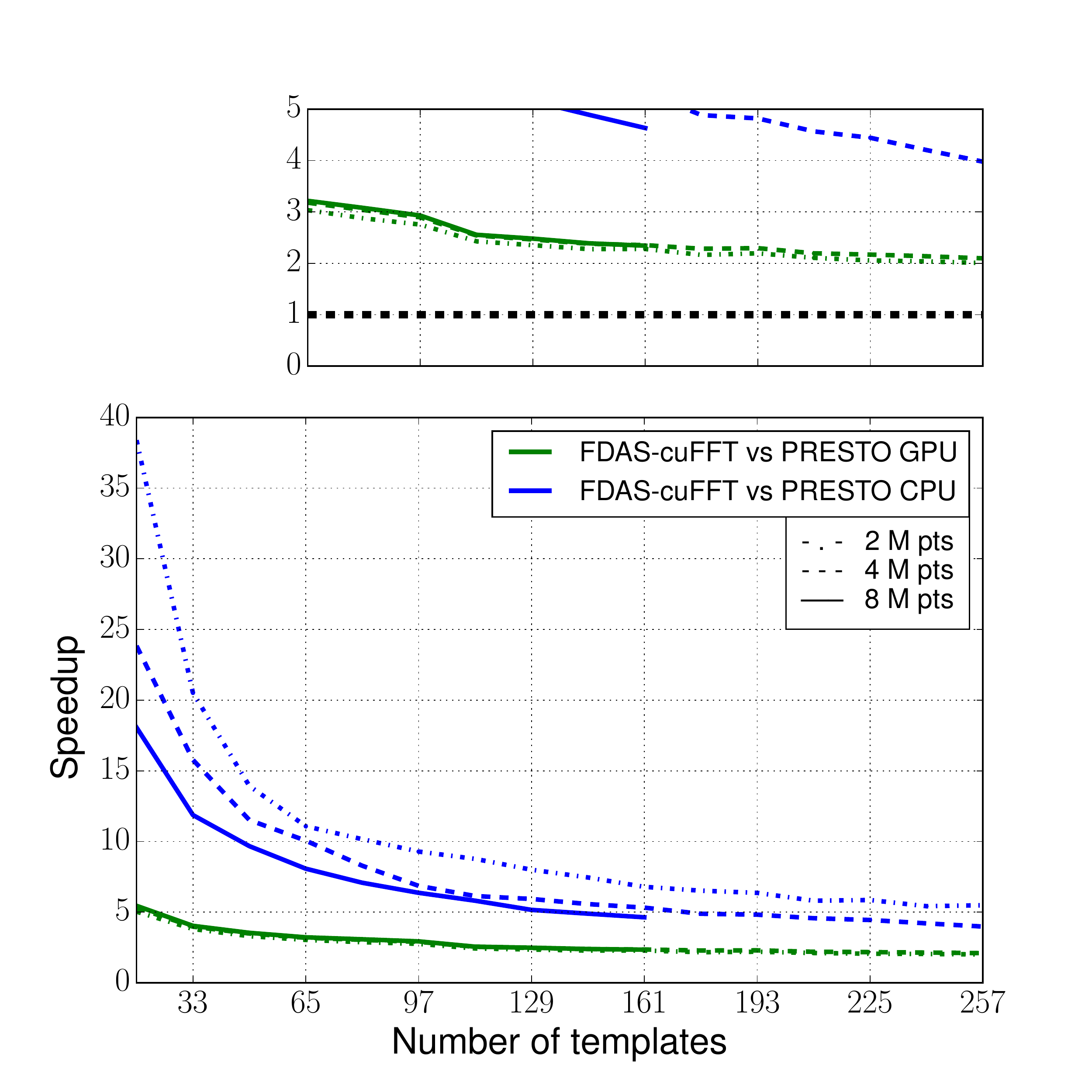}{0.5\textwidth}{(d)} \label{fig:f7d}
}
\caption{Comparisons between FDAS-cuFFT, PRESTO CPU and GPU on the NVIDIA M40 excluding the real FFT on CPU. (a), (b): Difference in execution time between codes,  (c), (d):  Speedup of FDAS-cuFFT compared to PRESTO CPU and GPU with the top zoomed to the y-axis at lower speedups and the black dashed line indicating where speedup=1.  Left: without interbinning, and right: with interbinning.}
\label{fig:f7}
\end{figure}

Examining the difference in execution time between the three codes without including the real FFT, we can see (fig.~\ref{fig:f7})  that the difference in performance between the GPU codes without the interbinning, shows a relatively small variation compared to its average, which indicates that the two codes have many similarities. However, PRESTO GPU has an overhead compared to FDAS-cuFFT, possibly due to the synchronous PCIe memory transfers and some additional indexing and other preparatory operations. We also observe that the FDAS correlation kernel provides a small performance increase between 32 and 96 templates. This difference is at the range of 50 \(m s\) with the longest input, although the speedup becomes very small for large numbers of templates. When interbinning is used, the FDAS implementation has a significant gain, maintaining twice the execution speed of PRESTO-GPU in the worst case, with an approximately linear increase in execution time difference. When comparing the GPU versions to the CPU we can also identify a proportional increase in the difference with the number of templates, which shows that the method improves in performance with size on the GPU.

Overall, if we exclude the real FFT calculations on the CPU, the basic FDAS-cuFFT implementation without the interbinning does not provide a significant improvement over the PRESTO-GPU version, although it does demonstrate an advantage for the case of real-time processing, by removing various overheads associated with PRESTO indexing calculations and PCIe memory transfers, and by optimising the correlation GPU kernel. The implementation of interbinning in FDAS offers a significant improvement over PRESTO GPU, with a minimum of double the execution speed.

\subsubsection{Comparisons between FDAS GPU versions with cuFFT and custom FFT }
\begin{figure}[htb!]
\gridline{\fig{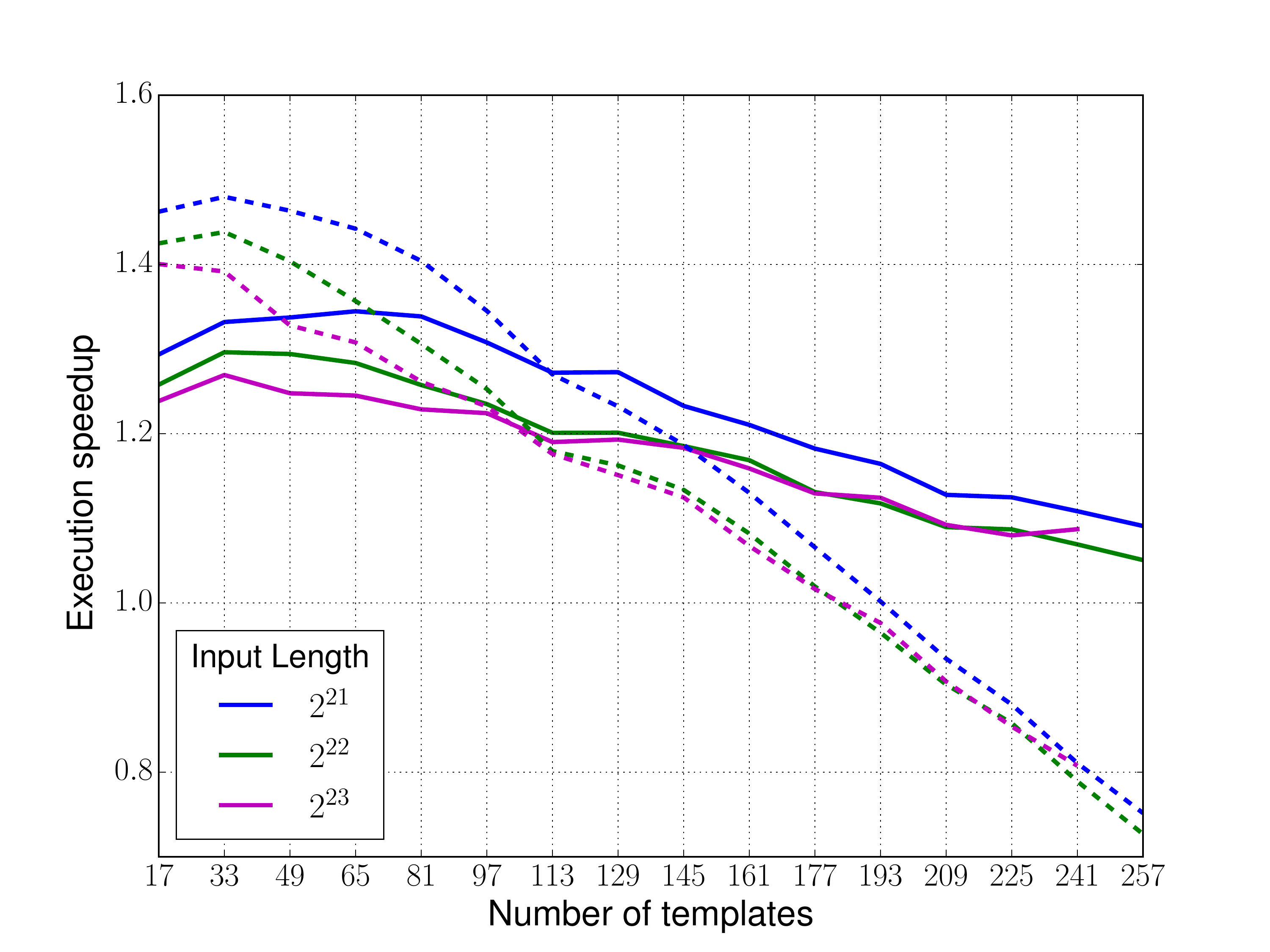}{0.5\textwidth}{(a)} \label{fig:f8a}
\fig{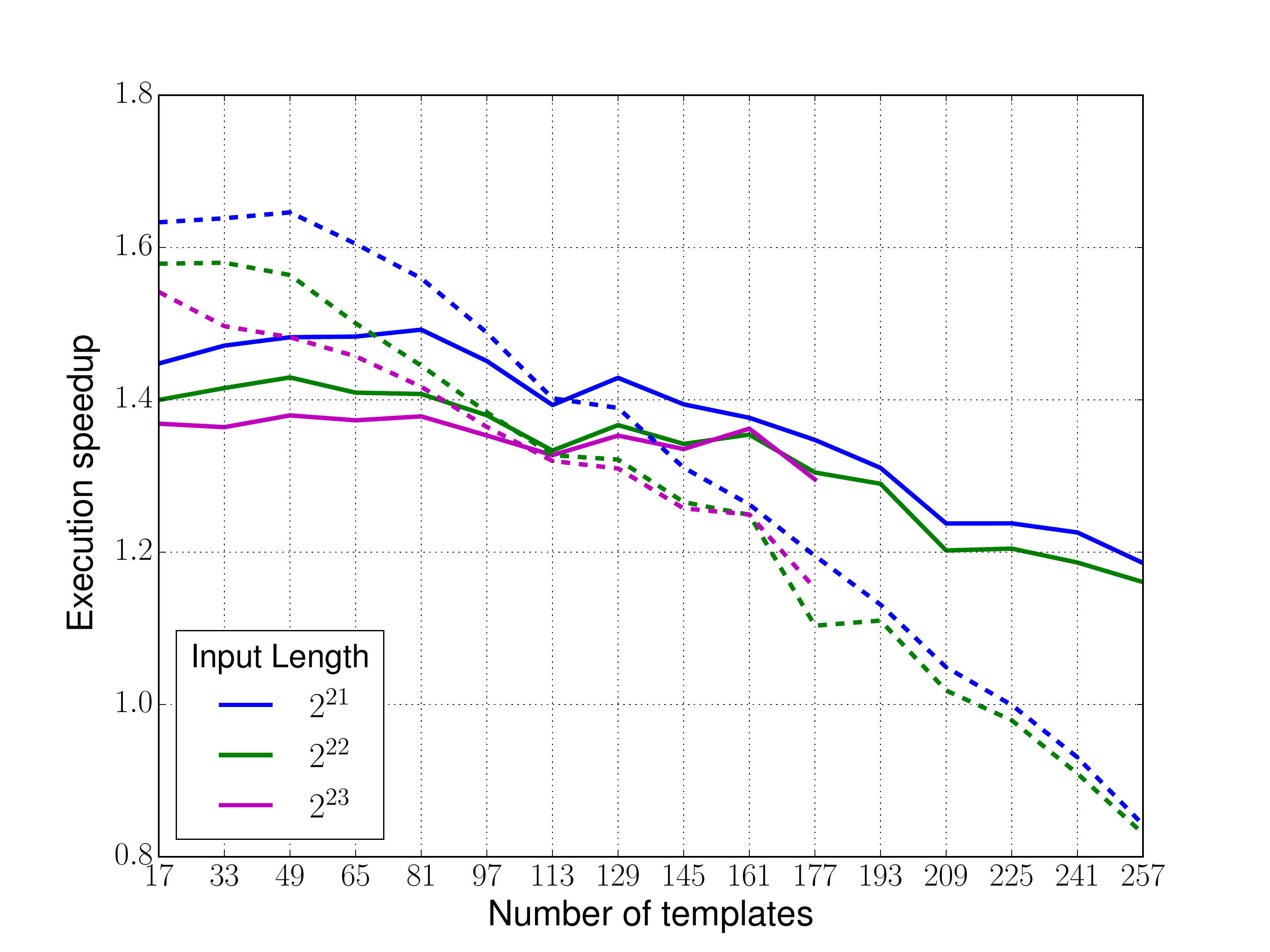}{0.5\textwidth}{(b)} \label{fig:f8b}
}
\gridline{
\fig{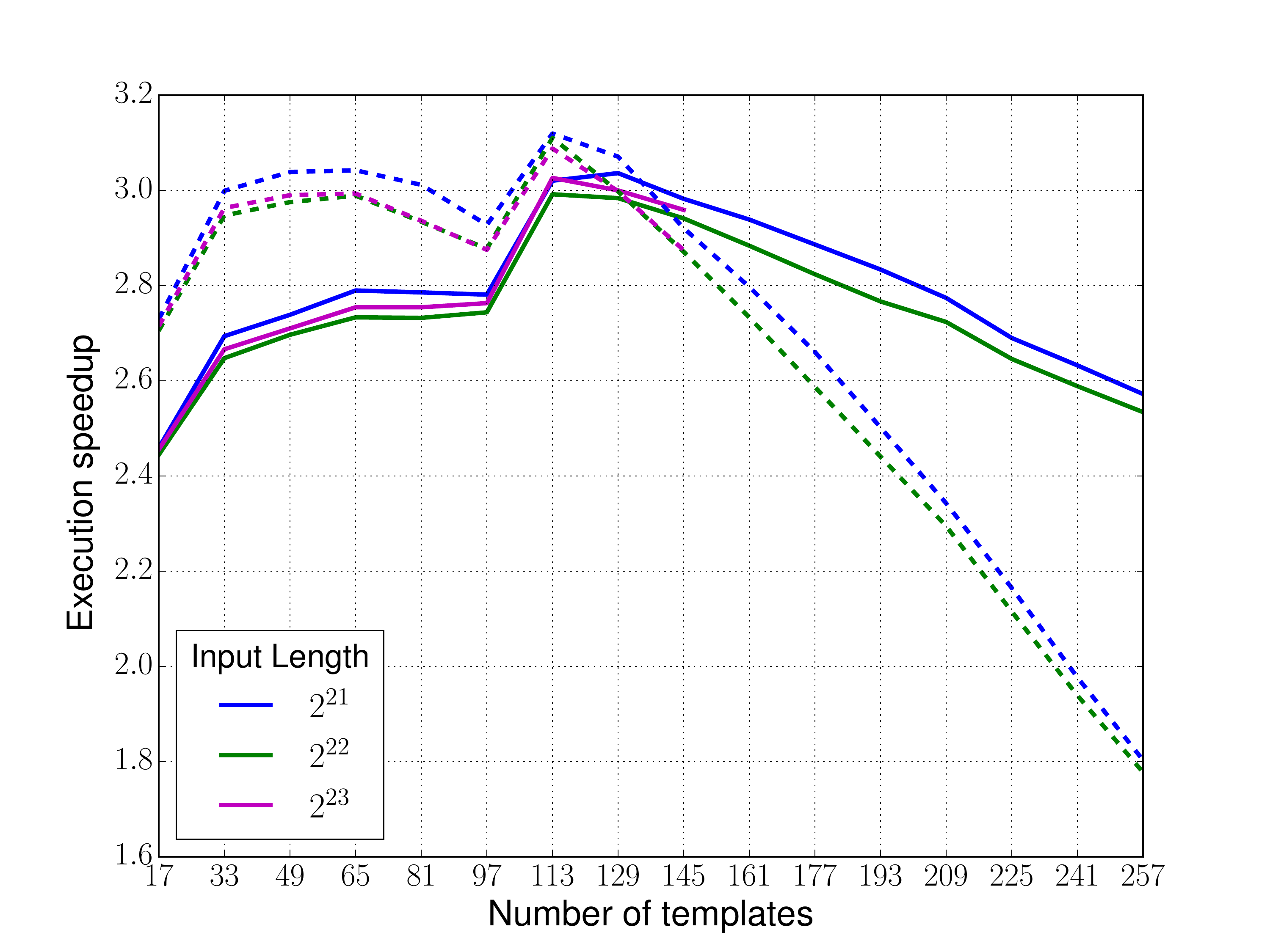}{0.5\textwidth}{(c)} \label{fig:f8c}
\fig{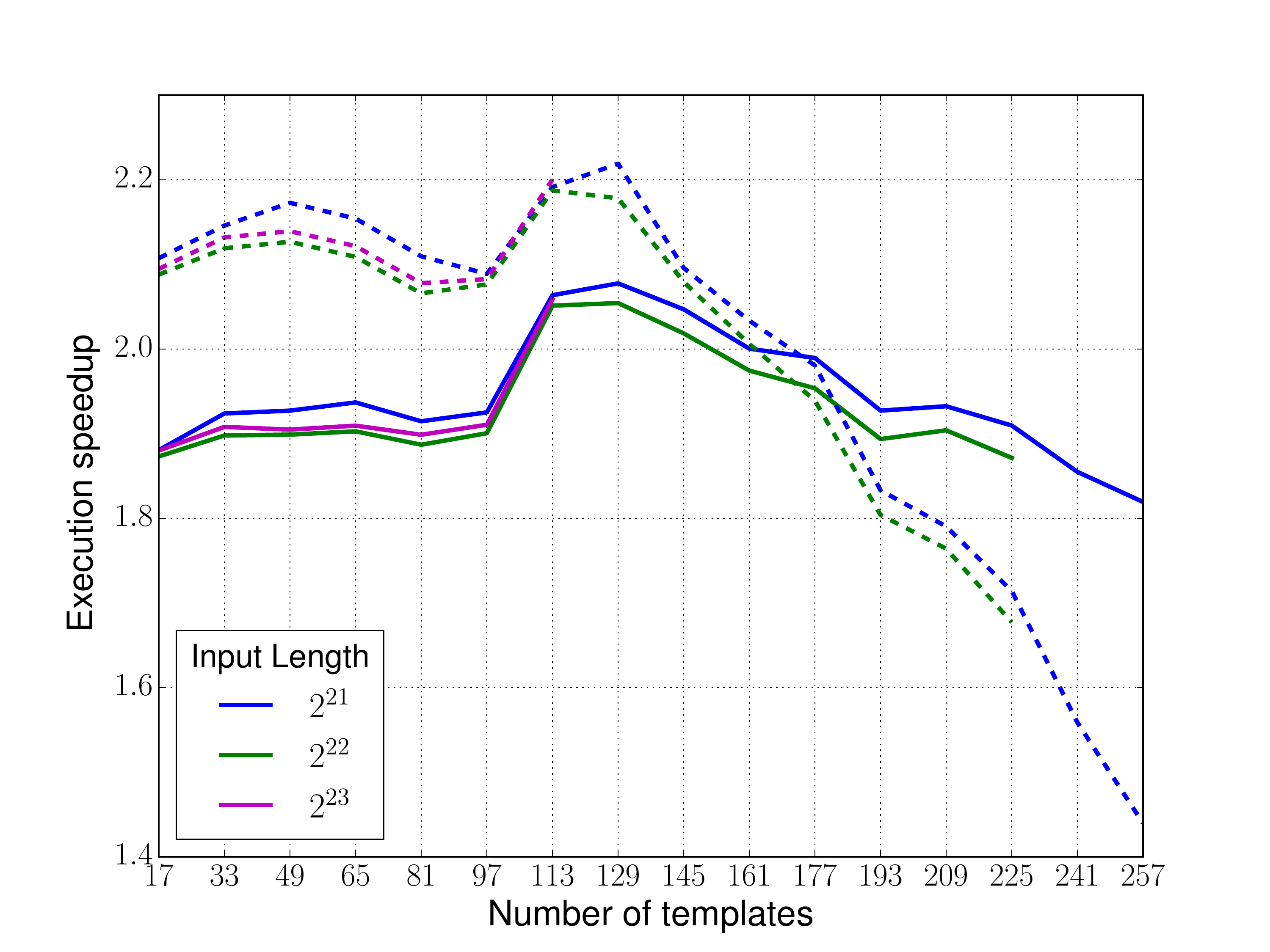}{0.5\textwidth}{(d)} \label{fig:f8d}
}

\gridline{
\fig{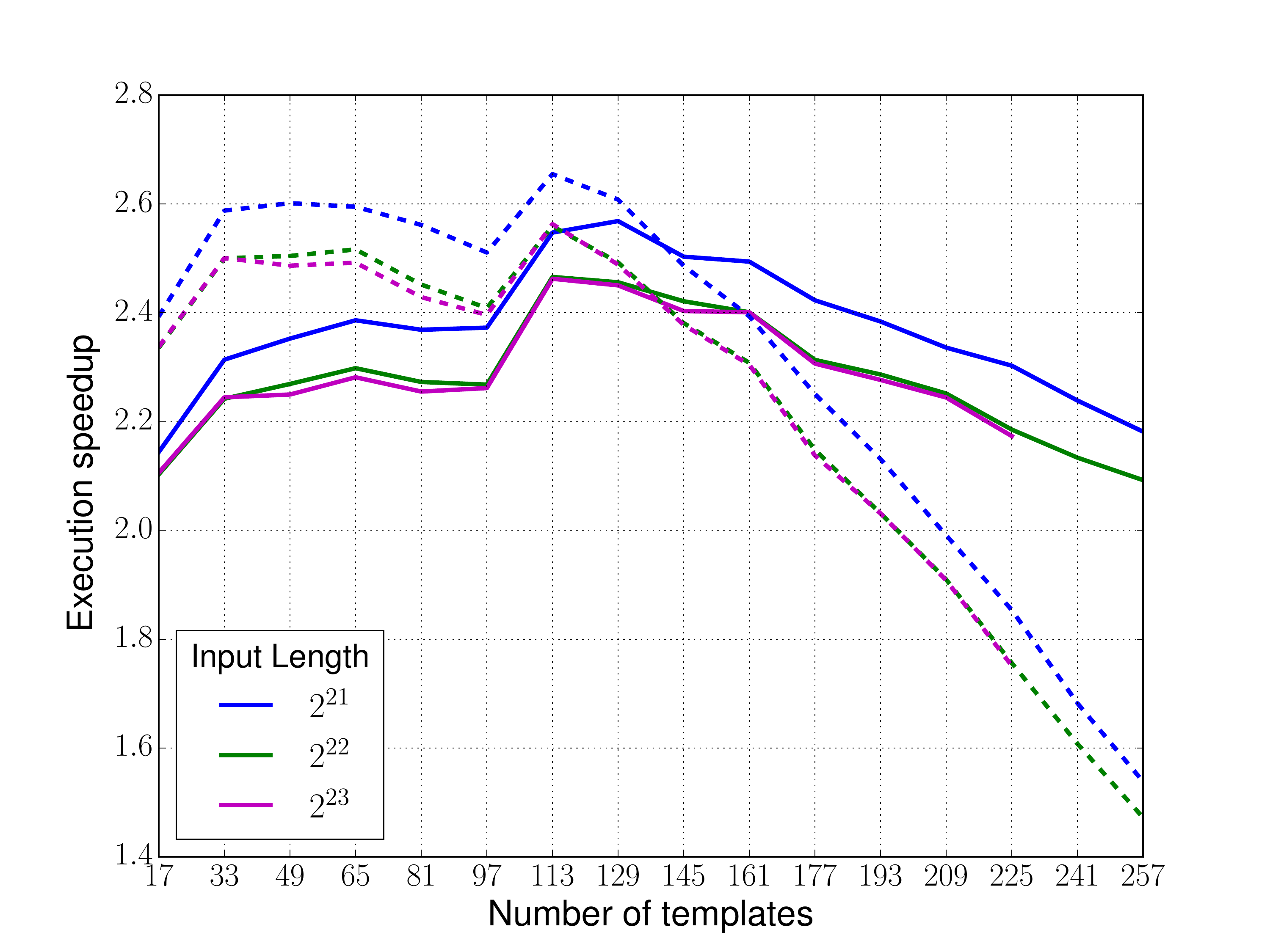}{0.5\textwidth}{(e)} \label{fig:f8e}
\fig{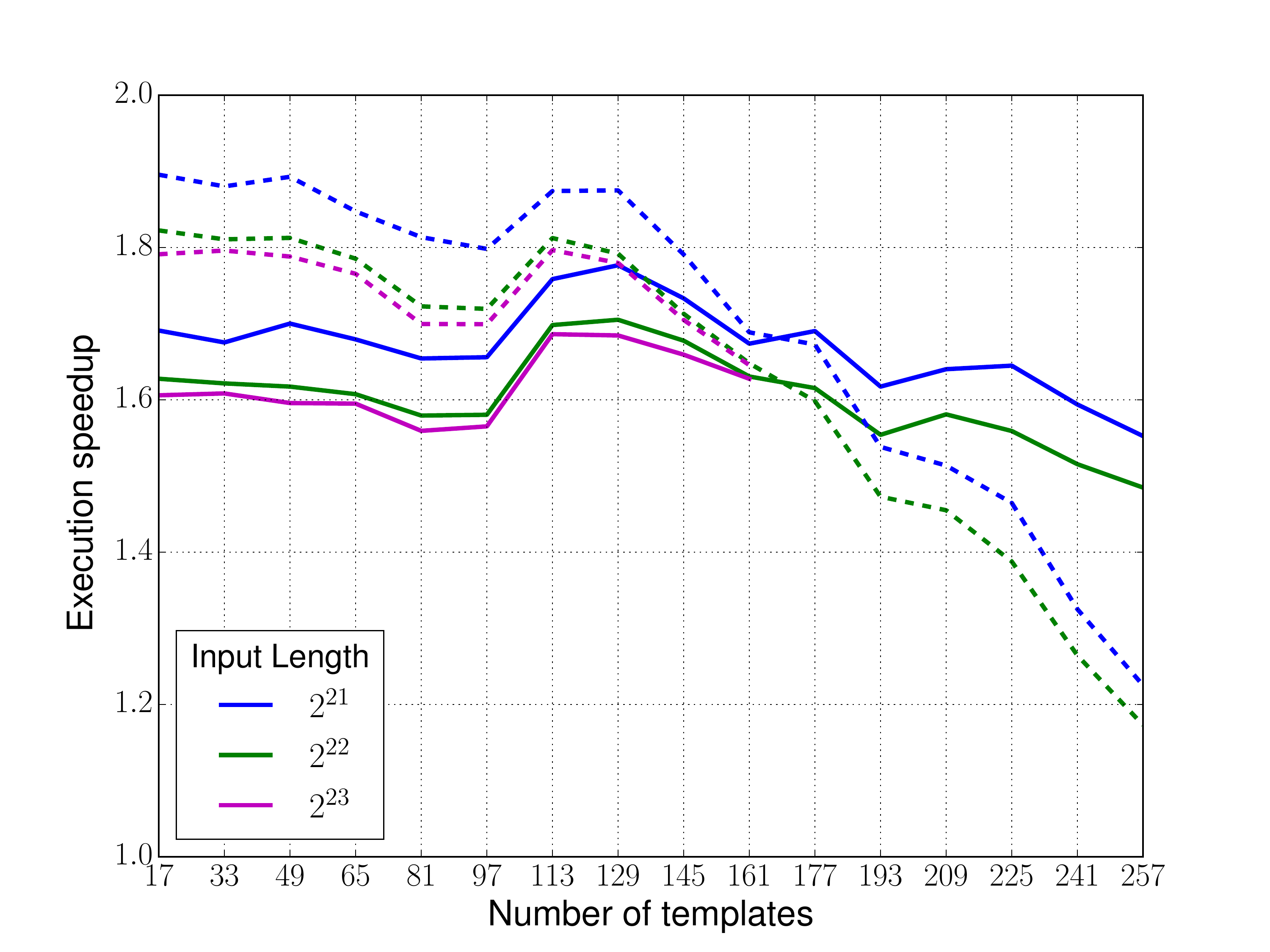}{0.5\textwidth}{(f)} \label{fig:f8f}
}
\caption{Comparison of FDAS with cuFFT and custom FFT on the NVIDIA Tesla K80 single GPU ((a), (b)), M60 single GPU ((c), (d)) and M40 ((e), (f)). Graphs on the left are without interbinning, and on the right with interbinning. The solid lines are for a template length of 1024 points, and the dashed lines for 512 points. Speed gains drop with increasing number of templates, but are maintained to a better level with a template length of 1024. }
\label{fig:f8}
\end{figure}

\begin{figure}[htb!]
\gridline{\fig{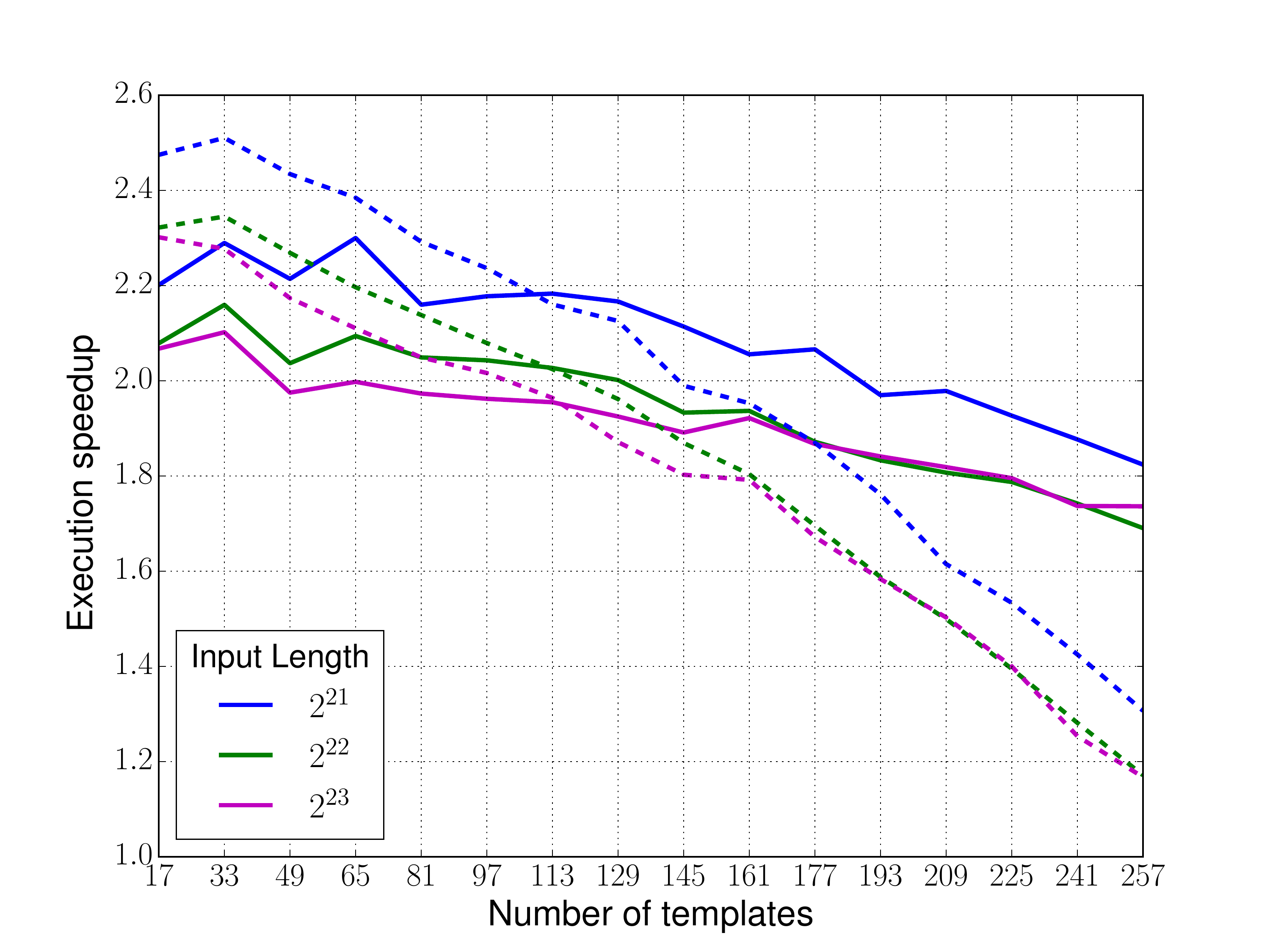}{0.5\textwidth}{(a)} \label{fig:f9a}
\fig{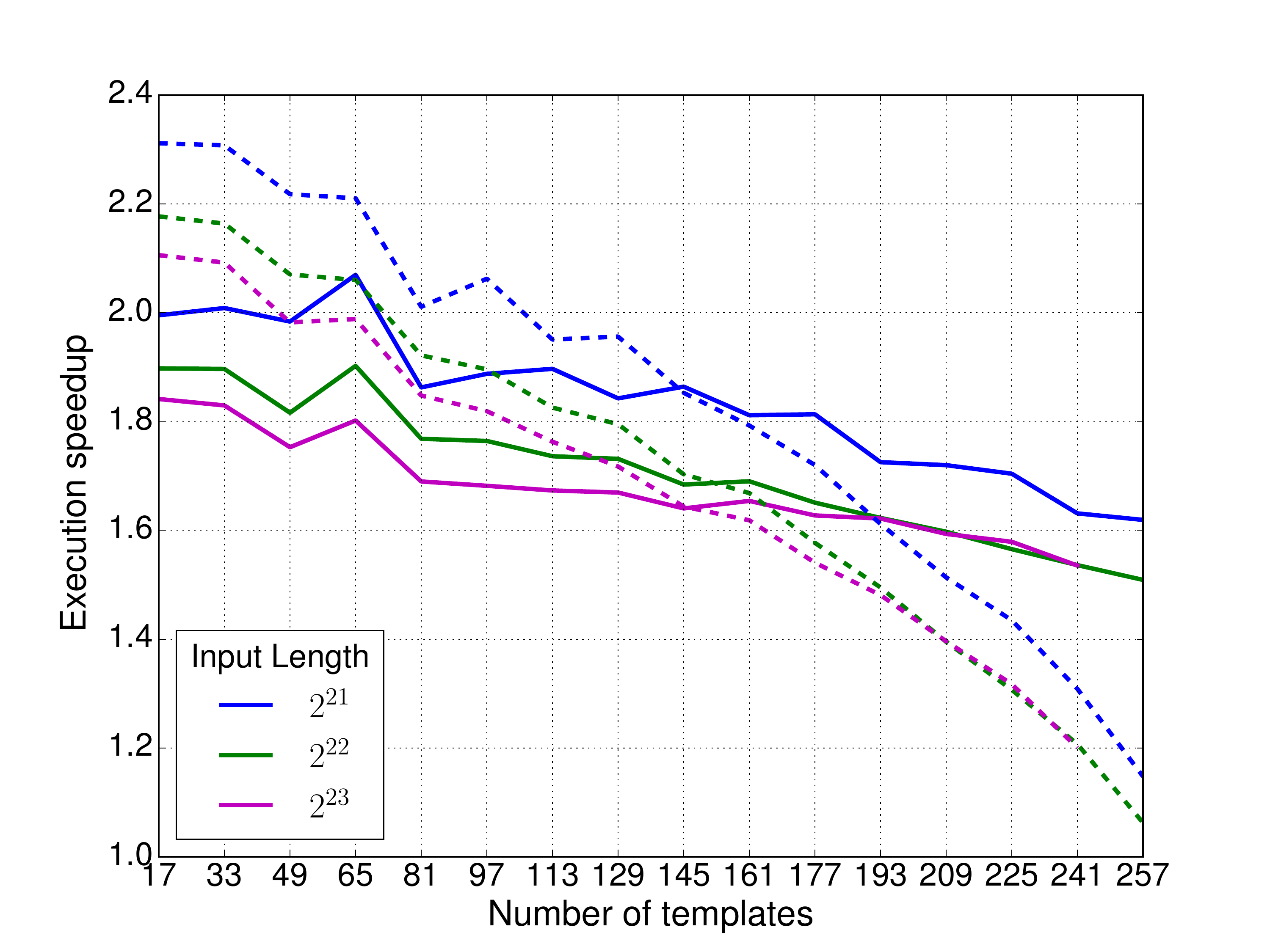}{0.5\textwidth}{(b)} \label{fig:f9b}
}
\caption{Speedup of FDAS with  custom FFT to FDAS with cuFFT on the NVIDIA P100 (a) without interbinning and (b) with interbinning. Without any optimisations we obtain better gains for smaller number of templates, and similar gains with higher number of templates, to that of the Maxwell architecture.}
\label{fig:f9}
\end{figure}

The cuFFT version is dominated by the Fourier transforms, whose speed is depended on the filter length, while the custom FFT version is restricted to a filter length of up to 1024 points. On the other hand, because of the overlap and save scheme, as the filter size decreases, the number of blocks to be convolved along the signal increases, and so does the number of the FFTs performed. For this reason we have chosen to run each code with its optimal filter sizes. Tests of both codes with varying filter sizes have shown that,  with the available testing hardware, the best size for the cuFFT version is 8192 points, while the custom FFT code had shown varying performance in different cases for the sizes of 512 and 1024 points, and we chose to use both of these in the comparison. Figure~\ref{fig:f8} shows the speedup of the custom FFT code with respect to the reference GPU code (FDAS cuFFT), against number of filter templates processed. Results are plotted separately for the cases of single bin and interbinning, and for the different GPU devices (see caption of fig.~\ref{fig:f8} for details). Solid lines indicate filter sizes of 512 points and dotted lines of 1024 points. The code has been tuned on each card for optimal performance. Preliminary measurements, without any specific optimisation or tuning are shown for the P100 card in figure~\ref{fig:f9}.

The speedup varies according to the number of templates and the filter size, but by choosing the optimal filter length, a speedup between 1.5 and 3.2 can be achieved on Maxwell architectures, depending on the number of templates.
 
We observe that in all cases, the speedup of FDAS with custom FFT against the cuFFT  follows a similar pattern. The lines for the 512  and 1024  long filters cross in a particular region of number of templates for each card. For numbers of templates below this region, filter sizes of 512 points perform better, while for higher template numbers, there is a continuous drop in speedup, which is worse for the 512-point filters. This indicates that with the increasing number of templates, the custom FFT code is becoming compute-bound, and the effect of the increased number of convolution blocks to be processed with the shorter filter lengths becomes evident. The worst case speedup is seen with the single GPU on the Kepler generation K80 card, however, performance gains are improved with interbinning. In contrast, with the Maxwell generation cards, there is an overall improvement in speedup in favour of the custom FFT, but this  is reduced when interbinning is used. Interbinning costs in execution time compared to non-interbinning on the K80 GPU are \(\sim \times 1.5\) for the cuFFT code and \(\times 1.2\) for the custom FFT, and \(\sim \times 1.1\) and \(\times 1.6\) on the Maxwell cards.  When using the single-GPU card M40, the speedup behaviour is very similar to the M60, but displays a lower level of speedup overall, which is consistent with the device memory bandwidth. In particular, the lowered memory bandwidth per GPU in the dual-GPU M60 card worsens the performance of the cuFFT code to 61\% and 79\% of that of the K80, and to 57\% and 58\% of that of the M40,  as it is largely memory bandwidth limited. However, this dual GPU  architecture is desirable because it can increase speed when running on both GPUs while keeping energy consumption lower than using a single GPU card. 

The custom FFT code clearly benefits from the Maxwell architecture. A key improvement in Maxwell from the Kepler architecture that enables this to happen is the shared memory capacity and bank size, that enable higher shared memory efficiency. Code profiling has shown that the custom FFT code is sensitive to shared memory efficiency, while the cuFFT code is almost entirely dependent on global memory bandwidth. In addition, the Maxwell processor has a dedicated shared memory of 96KB per multiprocessor, while the cache on the K80 is divided between L1 cache and shared memory. Much higher shared memory throughput is achieved on Maxwell (see Table ~\ref{table:t1}) , due to the fact that the K80 chip optimises this bandwidth only with the utilisation of an optional 8-byte bank size, and as a result, it performs worse for 32-bit floating point and integers. In particular the M60 card has a much lower peak memory bandwidth than the K80, and consequently, the relative performance of the cuFFT code is reduced.

Finally, on a P100 without any specific tuning, the execution time seems to scale with an approximately 75\% decrease, while the speedup between FDAS custom FFT and cuFFT, shown in figure~\ref{fig:f9}, is maintained at the level of 1.5, similar to the M40 for the highest number of templates. There is also considerable improvement with a filter size of 1024 for lower numbers of templates.

\subsection{Signal Recovery}

We assess the quality of the signal recovery performance with our custom FFT algorithm, which has achieved the fastest results. We have run a set of trials, using the PRESTO acceleration search code as a reference for signal recovery. Acceleration searches are known to cover a limited parameter space \citep{1991ApJ...368..504J,0004-637X-535-2-975,0004-637X-589-2-911,0004-637X-774-2-93,doi:10.1093/mnras/stv753}, and to ensure our comparison is applied within a space where PRESTO is known to be effective, we used simulated pulsars with typical orbital parameters of millisecond pulsars in  binary systems. 

The input signals were produced using the program ``fake" from the SIGPROC pulsar processing package. Using a pulsar period of 1.0, 3.0, 5.0 6.0, 7.0, and 10.0 \(ms\), and a short orbital period of 2.0 h in a circular orbit, we have modified the companion mass between 0.01 and 0.6 solar masses to represent  low to medium mass companions and obtain a range of acceleration values. The orbital phase was also varied to obtain negative accelerations. To ensure sufficient recovery of the first  signal harmonic, the input pulse level was kept relatively high, at 0.1 for a single pulse, and the signal was produced with a wide duty cycle of 45\%. The observation length is a constant 536 s with a sampling interval of 64 \(\mu s\). The upper edge channel frequency was set at 1550MHz with 4096 channels of width 0.075 MHz each, which is consistent with SKA design parameters. The peak Signal to Noise Ratio (SNR) produced by our custom GPU code was compared to that produced by PRESTO, and Fourier amplitudes were interpolated during the correlations in PRESTO, and interbinned after the correlations in FDAS custom FFT. As mentioned previously in section~\ref{sec:sec3}, due to a lack of a harmonic summing module, the SNR for both PRESTO and the FDAS custom FFT is measured only on the fundamental frequency without harmonic summing. Isolated  pulsars  were also produced with the same telescope parameters at each of the periods used for the accelerated pulsars. The peak SNR of the isolated and accelerated pulsars was measured. The measurements taken are the SNR peaks calculated immediately after the correlations in both codes. The SNR calculations in FDAS were done using the PRESTO SNR functions, which were embedded in FDAS custom FFT. Table~\ref{table:t2} lists the search results. The differences quoted are subtractions of the FDAS custom FFT result from the PRESTO result, and a negative sign indicates the respective PRESTO value is lower. \(\mathrm{\Delta z}\) indicates the difference in the frequency derivative bin where the peak was detected, and  \(\mathrm{\Delta r}\) the difference in frequency bins of the detected frequencies along the time series. As a duty cycle of 45\% is very high, we have also performed a limited number of searches in simulated pulsars with 10\% duty cycle, for a more realistic example. The observation files were created for a small subset of the existing dataset described above, using the same specifications except the duty cycle, which was modified to 10\%. The search results are displayed in table~\ref{table:t3}.
\begin{table}[htb!]
\caption{Comparative search results from PRESTO with Fourier interpolation (PR) and FDAS custom FFT with interbinning (FD) for 45\% duty cycle pulsars.}
\begin{center}
\begin{tabular}{ccccccc}
\hline

 \textbf{Period} ($\mathbf{ms}$) & \textbf{Accel. (PR)} ($\mathbf{m/s^2}$) & $\mathbf{\Delta z \, (bins)}$ &  $\mathbf{\Delta r \, (bins)}$  & \textbf{SNR (PR)} & \textbf{SNR (FD)} & \textbf{SNR isol. (PR)} \\ \hline
 1 & 12.48 &2  & 0 & 51.13 & 54.25 & 92.68 \\
  & -10.40 &0 & -0.5 & 47.01 & 52.26 & \\ 
  & 22.89 & 0 & 0 & 41.88 & 53.66 & \\ 
  &-25.00.&-2 & 0 & 36.75 & 52.74 & \\ 
 3 & 31.21 &0 & 0 & 69.61 & 78.08 & 81.95 \\ 
  & -31.20 &0 & 0 & 62.59 & 61.17 & \\
  & 43.69 &0 & 0 & 76.37 & 78.97 & \\
  & -49.94 &-2 & 0 & 58.59 & 59.65 & \\ 
 5 & 62.43 &0 & 0 & 58.68 & 55.70 & 65.70 \\
  & -72.83 &0 & -0.5 & 61.60 & 61.13 & \\
  & 83.24 &0 & 0 & 61.31 & 60.93 &\\
 6 & 87.41 &0 & 0 & 35.92 & 40.48 & 65.55 \\ 
  & -87.33 &0 & 0 & 37.26 & 37.74 & \\
  & 112.40 &0 & 0 & 37.08 & 48.98 & \\
  & -112.27 &0 & 0 & 39.37 & 44.20 & \\
 7 & 101.98 &0 & 0 & 51.78 & 59.00 & 68.66 \\
  & -101.88 &0 & 0 & 53.62 & 58.03 & \\
  & 116.56 &0 & 0 & 59.48 & 62.48 & \\
  & -116.43 &0 & 0 & 54.53 & 60.92 & \\
 10 & 83.25 &0 & 0 & 49.74 & 56.25 & 86.67 \\ \hline
\end{tabular}
\end{center}
\label{table:t2}
\end{table}%

\begin{table}[htb!]
	\caption{Comparative search results from PRESTO with Fourier interpolation (PR) and FDAS custom FFT with interbinning (FD) for 10\% duty cycle pulsars.}
	\begin{center}
		\begin{tabular}{ccccccc}
			\hline
			
	 \textbf{Period} ($\mathbf{ms}$) & \textbf{Accel. (PR)} ($\mathbf{m/s^2}$) & $\mathbf{\Delta z \, (bins)}$ &  $\mathbf{\Delta r \, (bins)}$  & \textbf{SNR (PR)} & \textbf{SNR (FD)} & \textbf{SNR isol. (PR)} \\ \hline
			1 & 10.40 & 0  & 0 & 31.39 & 33.65 & 43.73 \\
			3 & 31.21 &0 & 0 & 30.67 & 32.17 & 43.04\\ 
			5 & 62.43 &0 & 0 & 33.70 & 31.35 & 29.87 \\ 
			6 & 87.41 &0 & 0 & 23.01 & 24.52 & 30.63 \\
			7 & 87.41 &0 & 0.5 & 25.07 & 29.11 & 30.42 \\ 
			10 &83.24 &0 & 0 & 25.00 & 28.55 & 35.60 \\ \hline
		\end{tabular}
	\end{center}
	\label{table:t3}
\end{table}%

There are very few discrepancies in the frequency and frequency derivative bins where the signal peaks were detected, each corresponding  to a single f and z step, i.e. one half bin and 1 template in the computations. The discrepancies in SNR recovery vary, with the largest variations being due to a higher value detected by FDAS custom FFT. The cause of this result is unclear, and it does not appear to relate directly to the differences in the detected frequency and acceleration. Possible causes are the method of interbinning, which is an approximation to a 2-bin Fourier interpolation and is applied after the correlations, as well as the different FFT algorithms used, which are well known to vary in accuracy\footnote{Also see the FFTW website in page titled ``FFT Accuracy Benchmark Comments'' at \url{http://www.fftw.org/accuracy/comments.html}.} \citep{JohnsonFr08:burrus,Schatzman:1996}. Nevertheless, FDAS custom FFT is succeeding in signal detection  sufficiently close to the frequency and acceleration of interest, which we believe provides a proof of concept. More accurate results and further metrics can be also obtained in post-processing by fine-binning small parts of the $f-\dot{f}$ plane around the area of detection (see for example \citep{1538-3881-124-3-1788}). Finally, it should be noted that the single bin, interbinned, and Fourier interpolated data have different statistical properties that affect the SNR calculations. FDAS custom FFT at this stage does not have its own methods for signal significance calculations, and the subject was beyond the scope of this work. Future work however should address this issue by investigating the properties and effects of this interbinning scheme and by developing and implementing the appropriate methods for calculating the SNR. 

\section{Operational Scenario}  \label{sec:sec6}
To examine the potential of our algorithm for execution within a streaming pipeline with the latest GPU hardware, we run a set of acceleration searches on the Tesla P100 for input signal lengths of  $N = [2^{21}, 2^{22}, 2^{23}, 2^{24}]$ points. The signals were  processed with $m = [64, 96, 128, 196, 256]$ templates each, except of the longest signal, that could fit up to 242  and 254 templates on a single GPU. Using a sampling of $64 \mu s$, as proposed for the SKA, we derive the observation period for each time series, and define this as the real-time limit for the corresponding signal. We then calculate the number of independent time series that can be processed for each $(N, m)$ combination. Table~\ref{table:t4} lists the results.
\begin{table}
\caption{Number of independent time series of various numbers of samples processed with FDAS with custom FFT in real-time, for a constant sampling interval of 64 $\mu s$. }
\begin{center}
\begin{tabular}{cccc}
\hline
\textbf{\# samples in Tseries} & \textbf{\# templates} & \multicolumn{2}{c}{\textbf{ \# Tseries processed}} \\ \hline
& & \textbf{no interbins} & \textbf{with interbins} \\ \hline
& 64 & 41427 & 33893 \\
& 96 & 24900 & 19775 \\
$2^{20}$ & 128 & 18847 & 15277 \\
& 192 & 12436 & 9898 \\
& 256 & 8039 & 6620 \\ \hline 
& 64 & 41645 & 34086 \\
& 96 & 26801 & 21123 \\
$2^{21}$ & 128 & 20147 & 15568 \\
& 192 & 12216 & 9686 \\
& 256 & 8338 & 6757 \\ \hline 
& 64 & 41850 & 34322 \\
& 96 & 27614 & 21756 \\
$2^{22}$ & 128 & 20455 & 16079 \\
& 192 & 12629 & 10043 \\
& 256 & 8477 & 6867 \\ \hline 
& 64 & 42099 & 34196 \\
& 96 & 27826 & 21744 \\
$2^{23}$ & 128 &  20627 & 16199 \\
& 192 & 12807 & 10157 \\
& 256 & 8600 & 6954 \\ \hline 
& 64 & 42454 & 34393 \\
& 96 & 28040 & 21832 \\
$2^{24}$ & 128 & 20708 & 16275 \\
& 192 & 12828 & 10203\\
& 254 / 242 \footnote{Number of templates limited by the GPU memory capacity. Quoted numbers represent templates for: no interbins / with interbins}& 8741 & 7537 \\ \hline 

\end{tabular}
\end{center}
\label{table:t4}
\end{table}%

The numbers of time series  achieved are affected much more by the number of templates than the signal size, which is expected, since the signal size is directly related to the real-time limit, i.e. when the signal size is increased, the time limit is also increased. The lowest number achieved is over 6500. Although this scenario does not include other important and computationally intense parts of the processing, such as harmonic sums, these high numbers listed in table~\ref{table:t4} indicate that the matched filtering step has been accelerated to such an extent that there is enough execution overhead to allow for the rest of the processing.

\section{Conclusions} \label{sec:sec7}
We have presented two GPU implementations of the correlation technique for Fourier domain acceleration searches, one based on the use of the cuFFT library, and the other on a custom FFT computed entirely using on-chip GPU resources. Our custom FFT implementation provides significant execution speed gains over a wide range of template numbers, compared to our cuFFT code, as well as to the other existing GPU and OpenMP implementations. Overall, we found that we are able to achieve a speed increase of  at least 6 times from the existing CPU OpenMP version of the correlations in the PRESTO acceleration search, and at least 4 times that of the PRESTO GPU version.

We have seen the  performance advancing from the Kepler to the Maxwell GPU architecture, with a particular benefit in performance per energy consumption from dual GPU cards. The recently released Pascal architecture also features improvements in chip design which favour on chip resources, as well as shared memory bandwidth, all of which allow for the scalability of the algorithm to future architectures, something that was strongly indicated from our initial tests on the P100 card.

Our algorithm also appears to be capable of sufficiently accurate recovery of  accelerated signals in comparison to the PRESTO acceleration search code. The use of interbinning on the correlated Fourier amplitudes demonstrates a benefit in the algorithm's execution speed compared to Fourier interpolation, but the effect it has  on detection is not yet well understood. The particular scheme has not been characterised appropriately yet, and future analysis and investigations would be needed in order to determine and improve the accuracy of detection.

The utilisation of shared memory and registers for intermediate steps instead of global memory, can expand in the future to accommodate a harmonic summing, search, and candidate selection algorithm on the GPU, with the potential for big reductions in execution time and energy consumption for the acceleration search process, which would be particularly useful for real-time time-domain processing on data-intensive next generation instruments as the SKA.

\acknowledgments

This work is supported by a Leverhulme Trust Project Grant (ARTEMIS: Real-time discovery in Radio Astronomy). It has also received support from the members of the Oxford pulsar group, Christopher Williams and Jayanth Chennamangalam, as well as support from from Prof. Ben Stappers and the Time Domain Team, a collaboration between Oxford, Manchester and MPIfR Bonn, to design and build the SKA pulsar search capabilities. Scott M. Ransom is a Senior Fellow of the Canadian Institute for Advanced Research, and is partially funded by the National Science Foundation's Physics Frontiers Center award 1430284.

\vspace{5mm}
\software{PRESTO \citep{2011ascl.soft07017R}, 
	PRESTO 2 on GPU (\url{https://github.com/jintaoluo/presto2_on_gpu}), 
	SIGPROC \citep{Lorimer-2011}}

\bibliographystyle{aasjournal}
\bibliography{fdas}  
\listofchanges
\end{document}